\documentclass[aps,prb,preprint,groupedaddress,showpacs,showkeys]{revtex4}
\usepackage {graphicx}
\usepackage {tabularx}
\usepackage{color}
\usepackage {ulem}
\usepackage{amsmath}

\begin{document}

%\preprint{}

\title{Current-Density Functional Theory for the superconductor}



\author{Katsuhiko Higuchi}
%\thanks{}
\affiliation{Graduate School of Advanced Sciences of Matter, 
Hiroshima University, Higashi-Hiroshima 739-8527, Japan}

\author{Masahiko Higuchi}
%\thanks{}
\affiliation{Department of Physics, Faculty of Science, Shinshu University, 
Matsumoto 390-8621, Japan}


\date{\today}

\begin{abstract}
We present the current-density functional theory for the superconductor 
immersed in the magnetic field. The order parameter of the superconducting 
state, transverse component of the paramagnetic current-density, and 
electron density are chosen as basic variables that uniquely determine the 
equilibrium properties of the system. In order to construct this theory, the 
development of the approximate form of the exchange-correlation (xc) energy 
functional is indispensable as well as the derivation of the effective 
single-particle equation which makes it possible to reproduce the 
equilibrium densities mentioned above. The rigorous expression of the 
xc-energy functional is derived using the technique of the coupling-constant 
integration. Furthermore, the approximate form of the xc energy functional 
is proposed such that the energy gap resulting from the effective 
single-particle equation is consistent with the attractive interaction 
energy of the system. 
\end{abstract}


\maketitle

%\section{}
%\subsection{}
%\subsubsection{}

%****************************Sec I*****************************
\section{\label{secI}Introduction}
The superconductivity has been one of the main topics in the condensed 
matter physics. In order to clarify the properties of the superconducting 
state, the first-principles approaches which can evaluate the order 
parameter of the superconducting state (OPSS) quantitatively play an 
important role \cite{1,2,3,4,5,6,7,8,9,10,11,12,13,14,15,16,17,18,19,20,21} 
as well as the other approaches using the model 
Hamiltonian \cite{22}. One of the most popular first-principles theories is the 
density functional theory (DFT) \cite{23,24,25}, and its extension to the 
superconductor was proposed by Oliveira, Gross and Kohn (OGK) \cite{26,27}. 
Since it can predict not only the critical temperature of the superconductivity 
but also the properties of the OPSS, i.e., spatial distribution and spin 
symmetry for the OPSS, the OGK theory has been applied to a lot of superconductors \cite{1,2,3,4,5,6,7,8,9,10,11,12,13,14,15,16,17,18,19,20,21, 23,24,25,26,27}.

However, the OGK theory cannot in principle predict the critical 
current-density because the current-density is not chosen as a basic 
variable to be reproduced \cite{26}. From the viewpoint of the practical 
applications of superconductors, the critical current-density is also an 
important quantity as well as the critical temperature \cite{28,29}. The critical 
current-density is related to the critical magnetic field via the London 
equation, which is so-called Silsbee's rule \cite{30,31}. The larger the critical 
current-density is, the larger the critical magnetic field is \cite{30,31}. 
Therefore, if the critical current-density is needed, we have only to 
calculate the critical magnetic field. 

In order to develop the first-principles theory for calculating the critical 
magnetic field, we need to deal with the superconductor immersed in the 
external magnetic field. Specifically, we need to construct the 
current-density functional theory (CDFT) for the superconductor, in which 
the dependence of the OPSS on the external magnetic field can be predicted. 
We shall stress once again that it is necessary to construct the CDFT for 
the superconductor so as to predict the critical current-density that is a 
key quantity for the practical use of the superconductor.

The CDFT for the solids of the normal state has previously been proposed by 
Vignale and Rasolt \cite{32,33}, and on the basis of this theory, the CDFT for 
the superconductor has been developed by Kohn, Oliveira and Gross (KOG) 
\cite{27}. However, unfortunately, the KOG theory may be regarded as a prototype 
of the CDFT for the superconductor, and it should be improved and extended 
in the following points: 

(i) The OPSS is essentially the two-variable function concerning both spin 
and spatial coordinates, because the OPSS contains as a part the 
two-particle wave function resulting from the Bose-Einstein condensation 
(BEC) of the fermion system \cite{34,35}. In the previous work \cite{27}, the OPSS was 
treated only in the restricted form with the spatially local and 
spin-singlet one. This restricted form cannot describe the center of the 
gravity of two particles that form the Cooper pair, so that the vortex 
pattern of the mixed state of the type II superconductors cannot be 
described. The spatial and spin dependences of the OPSS should be treated as 
they are without any approximation.

(ii) In the proof of the Hohenberg-Kohn (HK) theorem which will be shown in 
Sec. 3, the technique of the constrained-search \cite{36,37,38,39,40,41,42,43} 
is used to 
avoid the difficulty of the assumption of the $v$-representability \cite{27,44}. 
Specifically, we will extend the extended constrained-search (ECS) theory 
\cite{39,40,41,42,43} to the case of the finite temperature \cite{25}, 
and apply it to the CDFT for the superconductor.

(iii) In the conventional CDFT, the electron density and paramagnetic 
current-density have been adopted as the basic variables that uniquely 
determine the ground-state properties or equilibrium properties of the 
system \cite{27,32,33,45,46,47}. However, they are not independent of each other. 
Specifically, the electron density and the longitudinal component of the 
paramagnetic current-density are related to each other via the equation of 
continuity. When performing the variational principle with respect to the 
basic variables, it is more convenient to choose as the basic variables the 
physical quantities that are independent of each other. Therefore, in this 
paper, the transverse component of the paramagnetic current-density is 
chosen as one of the basic variables instead of the whole components of the 
paramagnetic current-density.

(iv) In the development of the DFT-based theory or ECS-based theory, the 
following two issues are indispensable and are closely connected with each 
other. One is to derive the effective single-particle equation on the basis 
of the HK theorem, which is so-called the Kohn-Sham (KS) equation \cite{24,39}. 
The KS equation for the superconductor is of the type of the Bogoliubov-de 
Gennes (BdG) equation \cite{48}, which is hereafter called BdG-KS equation. The 
other is to develop the approximate form of the exchange-correlation (xc) 
energy functional that is contained in the BdG-KS equation. The former issue 
just corresponds to the above-mentioned (ii) and (iii). As for the latter 
issue, there have been no attempts in the previous work \cite{27}. In this paper, 
we will derive the rigorous expression for the xc energy functional, and 
present the approximate form which is consistent with the attractive 
interaction between electrons.

In this paper, we present the CDFT for the superconductor with particular 
emphasize on the above-mentioned four points. Especially, (iv) is the most 
essential point of whether the present theory works well or not. 
Organization of this paper is as follows. In Sec. II, the OPSS is reviewed 
on the basis of the definition of the superconductivity \cite{34}. In Sec. III, 
the ECS theory \cite{39,40,41,42,43} is extended to the case of the finite temperature. 
The HK theorem is proven without the assumption of the $v$-representability. 
Introducing the reference system in which the HK theorem holds, the BdG-KS 
equation is derived in the case of the finite temperature in Sec. IV. In 
Sec. V, the effective mean-field potentials that are contained in the BdG-KS 
equation are determined by requiring that the transverse component of the 
paramagnetic current-density, electron density and OPSS of the equilibrium 
state are reproduced in the reference system. The development of the xc 
energy functional is one of main topics in the present paper. Not only the 
rigorous expression but also the approximate forms of the xc energy 
functional are presented in Sec. VI. Finally, some concluding remarks are 
given in Sec. VII.
%
%**************************** SecII **********************************
\section{Order parameter of the superconducting state}
\label{secII}
The superconducting state can be regarded as the BEC of the fermion system 
\cite{34}. The BEC of the fermion system is defined by using the second-order 
reduced density matrix (RDM2) \cite{34,35}. According to this definition, when 
the BEC occurs in the system, one of eigenvalues for the RDM2 takes the 
value of $O(N)$ and the corresponding eigenfunction extends in terms of the 
coordinate of the center of the gravity but localizes in terms of the 
relative coordinate \cite{49}. In that case, it is shown that the following 
quantity takes the nonzero value \cite{50,51,52}:
\begin{equation}
\label{eq1}
\Delta ({\rm {\bf r}}_{2} \zeta_{2} ,{\rm {\bf r}}_{1} \zeta_{1} 
)=\left\langle {\hat{{\Delta }}({\rm {\bf r}}_{2} \zeta_{2} ,{\rm {\bf 
r}}_{1} \zeta_{1} )} \right\rangle 
\,\,\,\,\,\mbox{with}\,\,\,\,\,\hat{{\Delta }}({\rm {\bf r}}_{2} \zeta_{2} 
,{\rm {\bf r}}_{1} \zeta_{1} )=\psi ({\rm {\bf r}}_{1} \zeta_{1} )\psi 
({\rm {\bf r}}_{2} \zeta_{2} )
\end{equation}
where $\psi ({\rm {\bf r}}\zeta )$ is the field operator of electrons, and 
where ${\rm {\bf r}}$ and $\zeta $ are spatial and spin coordinates, 
respectively. The bracket in the RHS means the expectation value with 
respect to the ground state in the case of zero temperature or the 
statistical average in the case of the finite temperature. $\Delta ({\rm 
{\bf r}}_{2} \zeta_{2} ,{\rm {\bf r}}_{1} \zeta_{1} )$ is exactly the 
OPSS, and is coincident with the square root of the maximum eigenvalue 
multiplied by the corresponding eigenfunction for the RDM2 \cite{34,50}. 

Using the spin-polarized states $\varphi_{k_{i} \sigma_{i} } ({\rm {\bf 
r}})\chi_{\sigma_{i} } (\zeta )$, and the corresponding annihilation 
operator $C_{k_{i} \sigma_{i} } $, the field operator is rewritten as
\begin{equation}
\label{eq2}
\psi ({\rm {\bf r}}\zeta )=\sum\limits_{k_{i} } {\sum\limits_{\sigma_{i} 
=\uparrow ,\downarrow } {\varphi_{k_{i} \sigma_{i} } ({\rm {\bf 
r}})\chi_{\sigma_{i} } (\zeta )C_{k_{i} \sigma_{i} } } } ,
\end{equation}
where $\chi_{\sigma_{i} } (\zeta )$ is the spin function. Substituting Eq. 
(\ref{eq2}) into Eq. (\ref{eq1}), we have
\begin{eqnarray}
\label{eq3}
\Delta ({\rm {\bf r}}_{2} \zeta_{2} ,{\rm {\bf r}}_{1} \zeta_{1})
&=&{\displaystyle \frac{1}{2}}\sum\limits_{k_{i} } {\sum\limits_{k_{j} } {\left\langle 
{C_{k_{i} \uparrow } C_{k_{j} \uparrow } } \right\rangle \left| 
{{\begin{array}{*{20}c}
 {\varphi_{k_{i} \uparrow } ({\rm {\bf r}}_{1} )} \hfill & {\varphi_{k_{j} 
\uparrow } ({\rm {\bf r}}_{1} )} \hfill \\
 {\varphi_{k_{i} \uparrow } ({\rm {\bf r}}_{2} )} \hfill & {\varphi_{k_{j} 
\uparrow } ({\rm {\bf r}}_{2} )} \hfill \\
\end{array} }} 
\right|\chi_{\uparrow } (\zeta_{1} )} } \chi_{\uparrow } (\zeta_{2} ) \nonumber \\ 
&\!\!\!+&\!\!\!{\displaystyle \frac{1}{2}}\sum\limits_{k_{i} } {\sum\limits_{k_{j} } {\left\langle 
{C_{k_{i} \downarrow } C_{k_{j} \downarrow } } \right\rangle \left| 
{{\begin{array}{*{20}c}
 {\varphi_{k_{i} \downarrow } ({\rm {\bf r}}_{1} )} \hfill & {\varphi 
_{k_{j} \downarrow } ({\rm {\bf r}}_{1} )} \hfill \\
 {\varphi_{k_{i} \downarrow } ({\rm {\bf r}}_{2} )} \hfill & {\varphi 
_{k_{j} \downarrow } ({\rm {\bf r}}_{2} )} \hfill \\
\end{array} }} 
\right|\chi_{\downarrow } (\zeta_{1} )} } \chi_{\downarrow} (\zeta_{2} )  \\ 
&\!\!\!+&\!\!\!{\displaystyle \frac{1}{2}}\sum\limits_{k_{i} } {\sum\limits_{k_{j} } {\left\langle 
{C_{k_{i} \uparrow } C_{k_{j} \downarrow } } \right\rangle \left| 
{{\begin{array}{*{20}c}
 {\varphi_{k_{i} \uparrow } ({\rm {\bf r}}_{1} )} \hfill & {\varphi_{k_{j} 
\downarrow } ({\rm {\bf r}}_{1} )} \hfill \\
 {\varphi_{k_{i} \uparrow } ({\rm {\bf r}}_{2} )} \hfill & {\varphi_{k_{j} 
\downarrow } ({\rm {\bf r}}_{2} )} \hfill \\
\end{array} }} 
\right|\left\{ {\chi_{\uparrow } (\zeta_{1} )\chi _{\downarrow } (\zeta_{2} )
+\chi_{\downarrow } (\zeta_{1} )\chi _{\uparrow } (\zeta_{2} )} \right\}} } \nonumber \\ 
&\!\!\!+&\!\!\!{\displaystyle \frac{1}{2}}\sum\limits_{k_{i} } {\sum\limits_{k_{j} } {\left\langle 
{C_{k_{i} \uparrow } C_{k_{j} \downarrow } } \right\rangle \left\{ {\varphi 
_{k_{i} \uparrow } ({\rm {\bf r}}_{1} )\varphi_{k_{j} \downarrow } ({\rm 
{\bf r}}_{2} )+\varphi_{k_{j} \downarrow } ({\rm {\bf r}}_{1} )\varphi 
_{k_{i} \uparrow } ({\rm {\bf r}}_{2} )} \right\}\left| 
{{\begin{array}{*{20}c}
 {\chi_{\uparrow } (\zeta_{1} )} \hfill & {\chi_{\downarrow } (\zeta_{1} 
)} \hfill \\
 {\chi_{\uparrow } (\zeta_{2} )} \hfill & {\chi_{\downarrow } (\zeta_{2} 
)} \hfill \\
\end{array} }} \right|} } \nonumber 
\end{eqnarray}
The first three terms of Eq. (\ref{eq3}) are spin-triplet wave functions, and the 
fourth one is the spin-singlet wave function. Taking into account the facts 
that the eigenfunction and eigenvalue for the RDM2 mean the two-particle 
wave function and its occupation number in the BEC state, respectively 
\cite{35,50}, it is recognized that the OPSS reflects the spin and spatial 
symmetries of the pairing state in the superconductor. Therefore, it is 
possible to find which type of the superconducting state, e.g., spin-singlet 
or spin-triplet or their mixed one, occurs by checking the OPSS. 

Using the coordinate of center of gravity ${\rm {\bf R}}$ and relative 
coordinate ${\boldsymbol \rho }$, where 
${\rm {\bf R}}={\left( {{\rm {\bf r}}_{1} +{\rm {\bf r}}_{2} } \right)}/2$,
${\boldsymbol \rho }={\rm {\bf r}}_{1} -{\rm {\bf r}}_{2} $, 
the OPSS is rewritten as 
$\Delta \left( {{\rm {\bf R}} {\boldsymbol \rho } ;\zeta_{1} \zeta_{2} } \right)$ instead of 
$\Delta ({\rm {\bf r}}_{2} \zeta_{2} ,{\rm {\bf r}}_{1} \zeta_{1} )$. 
By observing the dependence of 
$\Delta \left( {{\rm {\bf R}} {\boldsymbol \rho } ;\zeta_{1} \zeta_{2} } \right)$ 
on ${\boldsymbol \rho }$, we can find the spatial broadening of the 
pairing states in the superconductor \cite{50}. This enables us to discuss how 
close the pairing two electrons are to the Bose particle, which would 
characterize the properties of the superconducting state. Furthermore, by 
observing the dependence of 
$\Delta \left( {{\rm {\bf R}} {\boldsymbol \rho } ;\zeta_{1} \zeta_{2} } \right)$ 
on ${\rm {\bf R}}$, spatial distribution pattern of 
the OPSS can be found explicitly, which would also characterize the 
superconducting state. For example, in the type II superconductor, the 
magnetic flux periodically penetrates the portion of the superconductor in 
which the OPSS disappears. The dependence of 
$\Delta \left( {{\rm {\bf R}} {\boldsymbol \rho } ;\zeta_{1} \zeta_{2} } \right)$
on ${\rm {\bf R}}$ shows the spatial structure and pattern of the magnetic flux directly. 
Thus, the OPSS should be expressed in the form of possessing both dependences of 
$ {\boldsymbol \rho }$ and ${\rm {\bf R}}$ \cite{35,50}.
%
%******************* SecIII *********************************************
\section{Hohenberg-Kohn Theorem}
\label{secIII}
%%%%
In the previous work \cite{27}, as mentioned in Sec. I, the $v$-representability of 
basic variables has been assumed in the proof of the HK theorem. The 
$v$-representability means the assumption where basic variables that determine 
the properties of the equilibrium state of the system have necessarily the 
corresponding external potentials \cite{44}. In this section, we present the HK 
theorem of the CDFT for the superconductor without this assumption. 
Specifically, we construct the finite-temperature version of the ECS theory 
\cite{39,40,41,42,43} that is suitable for the superconductor immersed in the external 
magnetic field.

%******************* SecIII-A *********************************************
\subsection{Hamiltonian}
Let us start with the Hamiltonian of the superconductor which is immersed in 
the beforehand-given electromagnetic fields ${\rm {\bf A}}_{given} ({\rm 
{\bf r}})$ and $v_{given} ({\rm {\bf r}})$. These fields should be 
determined in a self-consistent way to express the Meissner effect \cite{27}, 
which will be explained in the Sec. V. D. The Hamiltonian includes not only 
the Coulomb repulsive interaction but also the attractive interaction 
induced by some elementary excitation such as phonon. It is given by
\begin{eqnarray}
\label{eq4}
 \hat{{H}}&=&\int {\psi^{\dag }({\rm {\bf r}}\zeta )\left[ 
{{\frac{1}{2m}}\left\{ {{\rm {\bf p}}+e{\rm {\bf A}}_{given} ({\rm {\bf 
r}})} \right\}^{2}} \right]} \psi ({\rm {\bf r}}\zeta )d^{3}rd\zeta \nonumber \\ 
&\!\!+&\!\!{\frac{1}{2}}\!\int\!\!\!\!\!\int\! {\psi^{\dag }({\rm {\bf r}}\zeta )\psi^{\dag 
}({\rm {\bf {r}'}}{\zeta }'){\frac{e^{2}}{\left| {{\rm {\bf r}}-{\rm {\bf 
{r}'}}} \right|}}\psi ({\rm {\bf {r}'}}{\zeta }')\psi ({\rm {\bf r}}\zeta )} 
d^{3}rd\zeta d^{3}{r}'d{\zeta }' \nonumber \\ 
&\!\!+&\!\!{\frac{1}{2}}\!\int\!\!\!\!\!\int\!\! {\psi^{\dag }({\rm {\bf r}}_{1} \zeta_{1} 
)\psi^{\dag }({\rm {\bf r}}_{2} \zeta_{2} )w({\rm {\bf r}}_{1} \zeta_{1} 
{\rm {\bf r}}_{2} \zeta_{2} ;{\rm {\bf r}}_{3} \zeta_{3} {\rm {\bf r}}_{4} 
\zeta_{4} )} \psi ({\rm {\bf r}}_{3} \zeta_{3} )\psi ({\rm {\bf r}}_{4} 
\zeta_{4} )d^{3}r_{1} d\zeta_{1} d^{3}r_{2} d\zeta_{2} d^{3}r_{3} d\zeta 
_{3} d^{3}r_{4} d\zeta_{4} \nonumber \\ 
&\!\!+&\!\!\int {v_{given} ({\rm {\bf r}})\psi^{\dag }({\rm {\bf r}}\zeta )\psi 
({\rm {\bf r}}\zeta )} d^{3}rd\zeta. 
\end{eqnarray}
The second and third terms of the RHS denote the Coulomb repulsive 
interaction energy and attractive interaction energy via $w({\rm {\bf 
r}}_{1} \zeta_{1} {\rm {\bf r}}_{2} \zeta_{2} ;{\rm {\bf r}}_{3} \zeta 
_{3} {\rm {\bf r}}_{4} \zeta_{4} )$, respectively. Neglecting the surface 
integral at infinite distance in a usual way \cite{32,33,45,46}, and adopting the 
Coulomb gauge given by $\nabla \cdot {\rm {\bf A}}_{given} ({\rm {\bf 
r}})=0$, the Hamiltonian is rewritten as
\begin{equation}
\label{eq5}
\hat{{H}}=\hat{{T}}+\hat{{W}}_{1} +\hat{{W}}_{2} +\hat{{V}}_{1} 
+\hat{{V}}_{2} +\hat{{V}}_{3} ,
\end{equation}
with
\begin{eqnarray}
\label{eq6}
&&\hat{{T}}
=
\int {\psi^{\dag }({\rm {\bf r}}\zeta ){\frac{{\rm {\bf 
p}}^{2}}{2m}}} \psi ({\rm {\bf r}}\zeta )d^{3}rd\zeta , \\
\label{eq7}
&&\hat{{W}}_{1}
=
{\frac{1}{2}}\int\!\!\!\!\!\int\! {\psi^{\dag }({\rm {\bf r}}\zeta 
)\psi^{\dag }({\rm {\bf {r}'}}{\zeta }'){\frac{e^{2}}{\left| {{\rm {\bf 
r}}-{\rm {\bf {r}'}}} \right|}}\psi ({\rm {\bf {r}'}}{\zeta }')\psi ({\rm 
{\bf r}}\zeta )} d^{3}rd\zeta d^{3}{r}'d{\zeta }', \\
\label{eq8}
&&\hat{{W}}_{2}
\!=\!
{\frac{1}{2}}\!\int\!\!\!\!\!\int \!\!
{\psi^{\dag }({\rm {\bf r}}_{\!1} \zeta_{1})\psi^{\dag }({\rm {\bf r}}_{\!2} \zeta_{2})
w({\rm {\bf r}}_{\!1} \zeta_{1} {\rm {\bf r}}_{\!2} \zeta_{2} ;{\rm {\bf r}}_{\!3} \zeta_{3} {\rm 
{\bf r}}_{\!4} \zeta_{4} )
\psi ({\rm {\bf r}}_{\!3} \zeta_{3} )\psi ({\rm {\bf 
r}}_{\!4} \zeta_{4} )} d^{3}\!r_{\!1} d\zeta_{1} d^{3}\!r_{\!2} d\zeta_{2} 
d^{3}\!r_{\!3} d\zeta_{3} d^{3}\!r_{\!4} d\zeta_{4},\ \ \ \ \\
\label{eq9}
&&\hat{{V}}_{1} 
=
\int {v_{given} ({\rm {\bf r}})\hat{{n}}({\rm {\bf r}})} 
d^{3}r, \\
\label{eq10}
&&\hat{{V}}_{2} 
=
e\int {{\rm {\bf A}}_{given} ({\rm {\bf r}})\cdot {\rm {\bf 
\hat{{j}}}}_{p}^{(T)} ({\rm {\bf r}})d^{3}r} ,\\
\label{eq11}
&&\hat{{V}}_{3} 
=
{\frac{e^{2}}{2m}}\int {{\rm {\bf A}}_{given} ({\rm {\bf 
r}})^{2}\hat{{n}}({\rm {\bf r}})d^{3}r} ,
\end{eqnarray}
where the operators $\hat{{V}}_{1} $, $\hat{{V}}_{2} $ and $\hat{{V}}_{3} $ 
denote external potential energies, respectively, and where ${\rm {\bf 
\hat{{j}}}}_{p}^{(T)} ({\rm {\bf r}})$ and $\hat{{n}}({\rm {\bf r}})$ 
contained in the external energies are operators of the transverse component 
of the paramagnetic current-density and the electron density, respectively, 
which are given by
\begin{eqnarray}
\label{eq12}
{\rm {\bf \hat{{j}}}}_{p}^{(T)} ({\rm {\bf r}})
&=&{\frac{\hbar }{i4\pi m}}\int {\left[ {\left\{ {\nabla_{{\rm {\bf {r}'}}} 
\psi^{\dag }({\rm {\bf 
{r}'}}{\zeta }')\times \nabla_{{\rm {\bf {r}'}}} \psi ({\rm {\bf 
{r}'}}{\zeta }')} \right\}\times {\frac{{\rm {\bf r}}-{\rm {\bf 
{r}'}}}{\left| {{\rm {\bf r}}-{\rm {\bf {r}'}}} \right|^{3}}}} \right]} 
d^{3}{r}'d{\zeta }',\\
\label{eq13}
\hat{{n}}({\rm {\bf r}})
&=&
\int {\psi^{\dag }({\rm {\bf r}}\zeta )\psi ({\rm {\bf r}}\zeta )} d\zeta .
\end{eqnarray}

Concerning the concrete form of $w({\rm {\bf r}}_{1} \zeta_{1} {\rm {\bf 
r}}_{2} \zeta_{2} ;{\rm {\bf r}}_{3} \zeta_{3} {\rm {\bf r}}_{4} \zeta 
_{4} )$ in Eq. (\ref{eq8}), for example, the phonon-induced attractive interaction 
that gives the BCS reduced Hamiltonian \cite{53,54,55} is given by
\begin{eqnarray}
\label{eq14}
w({\rm {\bf r}}_{1} \zeta_{1} {\rm {\bf r}}_{2} \zeta_{2} ;{\rm {\bf 
r}}_{3} \zeta_{3} {\rm {\bf r}}_{4} \zeta_{4} )
&=&
2\chi_{\uparrow } (\zeta _{1} )\chi_{\downarrow } (\zeta_{2} )
\chi_{\downarrow } (\zeta_{3} )\chi _{\uparrow } (\zeta_{4} ) \nonumber \\ 
&\times& 
\sum\limits_{{\rm {\bf k}}} {\sum\limits_{{\rm {\bf {k}'}}(\ne {\rm 
{\bf k}})} {V_{{\rm {\bf k{k}'}}} \varphi_{{\rm {\bf {k}'}}\uparrow } ({\rm 
{\bf r}}_{1} )\varphi_{-{\rm {\bf {k}'}}\downarrow } ({\rm {\bf r}}_{2} 
)\varphi_{-{\rm {\bf k}}\downarrow }^{\ast } ({\rm {\bf r}}_{3} )\varphi 
_{{\rm {\bf k}}\uparrow }^{\ast } ({\rm {\bf r}}_{4} )} },
\end{eqnarray}
where $V_{{\rm {\bf k{k}'}}} $ is the attractive electron-electron 
interaction induced by the electron-phonon interaction \cite{53,54,55}, and where 
$\varphi_{{\rm {\bf k}}\sigma } ({\rm {\bf r}})\chi_{\sigma } (\zeta )$ is 
the electron state with the momentum ${\rm {\bf k}}$ and spin $\sigma $.

%******************* SecIII-B *********************************************
\subsection{Basic variables}
In the DFT-based or ECS-based theory, we first have to choose the basic 
variables that determine the properties of the equilibrium state of the 
system. The OPSS, which is given by Eq. (\ref{eq1}), should be chosen as one of 
basic variables because the superconducting properties are reflected in the 
OPSS as mentioned in Sec. II. Since this is generally a complex number, the 
complex conjugate $\Delta^{\ast }\left( {{\rm {\bf r}}\zeta ,{\rm {\bf 
{r}'}}{\zeta }'} \right)$ is also the basic variable to be chosen. In 
addition to these two quantities, the densities $\left\langle {{\rm {\bf 
\hat{{j}}}}_{p}^{(T)} \left( {{\rm {\bf r}}} \right)} \right\rangle $ and 
$\left\langle {\hat{{n}}\left( {{\rm {\bf r}}} \right)} \right\rangle $ 
should also be chosen as basic variables because these are coupled with the 
external potentials and will be shown to become basic variables inevitably 
in the proof of the HK theorem \cite{39}. Here, the meaning of the bracket is the 
same as that explained below Eq. (\ref{eq1}).

Thus, the basic variables of the CDFT for the superconductor are
\begin{eqnarray}
\label{eq15}
&&n({\rm {\bf r}})
=
\left\langle {\int {\psi^{\dag }({\rm {\bf r}}\zeta )\psi 
({\rm {\bf r}}\zeta )} d\zeta } \right\rangle , \\
\label{eq16}
&&{\rm {\bf j}}_{p}^{(T)} ({\rm {\bf r}})
=
{\frac{\hbar }{i4\pi m}}\left\langle {\int {\left[ {\left\{ {\nabla_{{\rm {\bf {r}'}}} 
\psi ^{\dag }({\rm {\bf {r}'}}{\zeta }')\times \nabla_{{\rm {\bf {r}'}}} \psi 
({\rm {\bf {r}'}}{\zeta }')} \right\}\times {\frac{{\rm {\bf r}}-{\rm {\bf 
{r}'}}}{\left| {{\rm {\bf r}}-{\rm {\bf {r}'}}} \right|^{3}}}} \right]} 
d^{3}{r}'d{\zeta }'} \right\rangle , \\
\label{eq17}
&&\Delta ({\rm {\bf r}}_{2} \zeta_{2} ,{\rm {\bf r}}_{1} \zeta_{1} )
=
\left\langle {\psi ({\rm {\bf r}}_{1} \zeta_{1} )\psi ({\rm {\bf r}}_{2} 
\zeta_{2} )} \right\rangle , \\
\label{eq18}
&&\Delta ({\rm {\bf r}}_{2} \zeta_{2} ,{\rm {\bf r}}_{1} \zeta_{1} )^{\ast }
=
\left\langle {\psi ({\rm {\bf r}}_{1} \zeta_{1} )\psi ({\rm {\bf r}}_{2} 
\zeta_{2} )} \right\rangle^{\ast }.
\end{eqnarray}

%******************* SecIII-C *********************************************
\subsection{Universal energy functional}
We define the universal energy functional as
\begin{equation}
\label{eq19}
F\left[ n,{\rm {\bf j}}_{p}^{(T)} ,\Delta ,\Delta^{\ast } \right]
=\mathop{\mbox{Min}}\limits_{\hat{\rho}\to n,{\rm {\bf j}}_{p}^{(T)},\Delta ,\Delta^{\ast }} 
\mbox{Tr}\left\{ {\hat{{\rho }}\left( 
{\hat{{T}}+\hat{{W}}_{1} +\hat{{W}}_{2} } \right)+{\frac{1}{\beta 
}}\hat{{\rho }}\ln \hat{{\rho }}} \right\},
\end{equation}
where $\beta =1/({k_{B} T})$, and where 
$k_{B} $ and $T$ are Boltzmann factor and temperature of the system, 
respectively. The RHS of Eq. (\ref{eq19}) means that the minimization of the 
statistical average of energies and entropy operators that are independent 
of the external potentials is searched by varying the statistical operator 
$\hat{{\rho }}$ within the set of $\hat{{\rho }}$'s which yield prescribed 
following densities
\begin{eqnarray}
\label{eq20}
&&n({\rm {\bf r}})
=
\mbox{Tr}\left\{ {\hat{{\rho }}\ \hat{{n}}({\rm {\bf r}})} 
\right\}\,, \\
\label{eq21}
&&{\rm {\bf j}}_{p}^{(T)} ({\rm {\bf r}})
=
\mbox{Tr}\left\{ {\hat{{\rho }}\ {\rm 
{\bf \hat{{j}}}}_{p}^{(T)} ({\rm {\bf r}})} \right\}, \\
\label{eq22}
&&\Delta ({\rm {\bf r}}\zeta ,{\rm {\bf {r}'}}{\zeta }')
=
\mbox{Tr}\left\{ 
{\hat{{\rho }}\ \hat{{\Delta }}({\rm {\bf r}}\zeta ,{\rm {\bf {r}'}}{\zeta 
}')} \right\}, \\
\label{eq23}
&&\Delta^{\ast }({\rm {\bf r}}\zeta ,{\rm {\bf {r}'}}{\zeta}')
=
\mbox{Tr}\left\{ {\hat{{\rho }}\ \hat{{\Delta }}^{\dag }({\rm {\bf 
r}}\zeta ,{\rm {\bf {r}'}}{\zeta }')} \right\}.
\end{eqnarray}
Using this universal energy functional, the HK theorem of the CDFT for the 
superconductor can be proved, which is shown in the next subsection.

%******************* SecIII-D *********************************************
\subsection{Hohenberg-Kohn theorem}
The HK theorem consists of two kinds of theorems. One is the variational 
principle with respect to densities given by Eqs. (\ref{eq20}) - (\ref{eq23}), and the other 
is the one-to-one correspondence between the correct statistical operator 
$\hat{{\rho }}_{0} $ and the densities of the equilibrium state. These 
theorems are the fundament of the CDFT for the superconductor. In what 
follows, we shall give the proof of the HK theorem. 

Gibbs's variational principle is written as \cite{25}
\begin{equation}
\label{eq24}
J_{0} =\mathop{\mbox{Min}}\limits_{\hat{{\rho }}} J\left[ {\hat{{\rho }}} 
\right]=J\left[ {\hat{{\rho }}_{0} } \right],
\end{equation}
with
\begin{equation}
\label{eq25}
J\left[ {\hat{{\rho }}} \right]=\mbox{Tr}\left\{ {\hat{{\rho }}\left( 
{\hat{{H}}-\mu \hat{{N}}} \right)+{\frac{1}{\beta }}\hat{{\rho }}\ln 
\hat{{\rho }}} \right\},
\end{equation}
where $\hat{{N}}$ is the operator of the electron number, and where $\mu $ 
is the chemical potential of the system. In Eq. (\ref{eq24}), $J_{0} $ is the grand 
potential of the equilibrium state, and the minimizing statistical operator 
corresponds to the correct one, which is given by
\begin{equation}
\label{eq26}
\hat{{\rho }}_{0} ={\frac{e^{-\beta \left( {\hat{{H}}-\mu \hat{{N}}} 
\right)}}{\Xi }},\,\,\,
\end{equation}
with $\Xi =\mbox{Tr}\left( {e^{-\beta \left( {\hat{{H}}-\mu \hat{{N}}} 
\right)}} \right)$. This variational principles can be rewritten using 
two-step variations such that
\begin{equation}
\label{eq27}
J_{0} =\mathop{\mbox{Min}}\limits_{\hat{{\rho }}} J\left[ {\hat{{\rho }}} 
\right]=\mathop{\mbox{Min}}\limits_{n,{\rm {\bf j}}_{p}^{(T)} ,\Delta ,\Delta 
^{\ast }} \left\{ {\mathop{\mbox{\!\!\!\!\!\!\!Min}}\limits_{\hat{{\rho }}\to n,{\rm {\bf 
j}}_{p}^{(T)} ,\Delta ,\Delta^{\ast }} J\left[ {\hat{{\rho }}} \right]} 
\right\}.
\end{equation}
Substituting Eqs. (\ref{eq25}) and (\ref{eq5}) into Eq. (\ref{eq27}), and using Eqs. (\ref{eq9}) -- (\ref{eq11}) 
together with Eqs. (\ref{eq20}) and (\ref{eq21}), we get
\begin{eqnarray}
\label{eq28}
 J_{0} =\mathop{\mbox{Min}}\limits_{n,{\rm {\bf j}}_{p}^{(T)} ,\Delta ,\Delta 
^{\ast }} \left\{ {F\left[ {n,{\rm {\bf j}}_{p}^{(T)} ,\Delta ,\Delta^{\ast 
}} \right]+\int {\left\{ {v_{given} ({\rm {\bf r}})-\mu } \right\}n({\rm 
{\bf r}})} d^{3}r} \right.\nonumber  \\ 
 \left. {\,\,\,\,\,\,\,\,\,\,\,\,\,\,\,\,\,+e\int {{\rm {\bf A}}_{given} 
({\rm {\bf r}})\cdot {\rm {\bf j}}_{p}^{(T)} ({\rm {\bf r}})d^{3}r+} 
{\displaystyle\frac{e^{2}}{2m}}\int {{\rm {\bf A}}_{given} ({\rm {\bf r}})^{2}n({\rm {\bf 
r}})d^{3}r} } \right\} , 
\end{eqnarray}
where Eq. (\ref{eq19}) is used. If we define the following functional
\begin{eqnarray}
\label{eq29}
 J^{v_{given} -\mu ,\,{\rm {\bf A}}_{given} }\left[ {n,{\rm {\bf 
j}}_{p}^{(T)} ,\Delta ,\Delta^{\ast }} \right]=F\left[ {n,{\rm {\bf 
j}}_{p}^{(T)} ,\Delta ,\Delta^{\ast }} \right]+\int {\left\{ {v_{given} 
({\rm {\bf r}})-\mu } \right\}n({\rm {\bf r}})} d^{3}r \nonumber \\ 
 +e\int {{\rm {\bf A}}_{given} ({\rm {\bf r}})\cdot {\rm {\bf j}}_{p}^{(T)} 
({\rm {\bf r}})d^{3}r+} {\displaystyle\frac{e^{2}}{2m}}\int {{\rm {\bf A}}_{given} 
({\rm {\bf r}})^{2}n({\rm {\bf r}})d^{3}r} ,
\end{eqnarray}
then Eq. (\ref{eq28}) is rewritten as
\begin{equation}
\label{eq30}
J_{0} =\mathop{\mbox{Min}}\limits_{n,{\rm {\bf j}}_{p}^{(T)} ,\Delta ,\Delta^{\ast 
}} J^{v_{given} -\mu ,\,{\rm {\bf A}}_{given} }\left[ {n,{\rm {\bf 
j}}_{p}^{(T)} ,\Delta ,\Delta^{\ast }} \right].
\end{equation}
Compared the second equality of the RHS of Eq. (\ref{eq27}) with Eq. (\ref{eq30}), the 
functional $J^{v_{given} -\mu ,\,{\rm {\bf A}}_{given} }\left[ {n,{\rm {\bf 
j}}_{p}^{(T)} ,\Delta ,\Delta^{\ast }} \right]$ is written as
\begin{equation}
\label{eq31}
J^{v_{given} -\mu ,\,{\rm {\bf A}}_{given} }\left[ {n,{\rm {\bf 
j}}_{p}^{(T)} ,\Delta ,\Delta^{\ast }} 
\right]=\mathop{\mbox{Min}}\limits_{\hat{{\rho }}\to n,{\rm {\bf j}}_{p}^{(T)} 
,\Delta ,\Delta^{\ast }} J\left[ {\hat{{\rho }}} \right].
\end{equation}
Equation (\ref{eq31}) means that the value of $J^{v_{given} -\mu ,\,{\rm {\bf 
A}}_{given} }\left[ {n,{\rm {\bf j}}_{p}^{(T)} ,\Delta ,\Delta^{\ast }} 
\right]$ corresponds to the grand potential at the minimum point within the 
restricted set of density matrices that yield the prescribed denisties 
$(n,{\rm {\bf j}}_{p}^{(T)} ,\Delta ,\Delta^{\ast })$. Therefore, Eq. (\ref{eq30}) 
means that the global minimum point is searched within the set of local 
minimum points searched by Eq. (\ref{eq31}). Since such a global minimum point gives 
the correct grand potential $J_{0} $, and using Eq. (\ref{eq24}), the densities 
$(n,{\rm {\bf j}}_{p}^{(T)} ,\Delta ,\Delta^{\ast })$ that are found via 
Eq. (\ref{eq30}) correspond to those that are calculated from the correct 
statistical operator $\hat{{\rho }}_{0} $. They are exactly the electron 
density, transverse component of the paramagnetic current-density and OPSS 
for the equilibrium state of the system, which are hereafter denoted as 
$n_{0} ,\,\,{\rm {\bf j}}_{p0}^{(T)} ,\,\,\Delta_{0} $ and $\Delta_{0} 
^{\ast }$, respectively. Thus, Eq. (\ref{eq30}) represents the variational principle 
with respect to the electron density, transverse component of the 
paramagnetic current density and OPSS. The results are summarized as follows:
\begin{equation}
\label{eq32}
J_{0} =\mathop{\mbox{Min}}\limits_{n,{\rm {\bf j}}_{p}^{(T)} ,\Delta ,\Delta^{\ast 
}} J^{v_{given} -\mu ,\,{\rm {\bf A}}_{given} }\left[ {n,{\rm {\bf 
j}}_{p}^{(T)} ,\Delta ,\Delta^{\ast }} \right]=J^{v_{given} -\mu ,\,{\rm 
{\bf A}}_{given} }\left[ {n_{0} ,{\rm {\bf j}}_{p0}^{(T)} ,\Delta_{0} 
,\Delta_{0}^{\ast }} \right],
\end{equation}
with
\begin{eqnarray}
\label{eq33}
&&n_{0} ({\rm {\bf r}})
=
\mbox{Tr}\left\{ {\hat{{\rho }}_{0} \,\hat{{n}}({\rm 
{\bf r}})} \right\}\,, \\
\label{eq34}
&&{\rm {\bf j}}_{p0}^{(T)} ({\rm {\bf r}})
=
\mbox{Tr}\left\{ {\hat{{\rho }}_{0} \,
{\rm {\bf \hat{{j}}}}_{p}^{(T)} ({\rm {\bf r}})} \right\},\\
\label{eq35}
&&\Delta_{0} ({\rm {\bf r}}\zeta ,{\rm {\bf {r}'}}{\zeta }')
=
\mbox{Tr}\left\{ 
{\hat{{\rho }}_{0} \,\hat{{\Delta }}({\rm {\bf r}}\zeta ,{\rm {\bf 
{r}'}}{\zeta }')} \right\}, \\
\label{eq36}
&&\Delta_{0}^{\ast }({\rm {\bf r}}\zeta ,{\rm {\bf {r}'}}{\zeta }')
=
\mbox{Tr}\left\{ {\hat{{\rho }}_{0} \,\hat{{\Delta }}^{\dag }({\rm {\bf 
r}}\zeta ,{\rm {\bf {r}'}}{\zeta }')} \right\}.
\end{eqnarray}

Next, we shall give the proof of the one-to-one correspondence between the 
correct statistical operator $\hat{{\rho }}_{0} $ and equilibrium densities 
$(n_{0} ,\,\,{\rm {\bf j}}_{p0}^{(T)} ,\,\,\Delta_{0} ,\,\,\Delta_{0} ^{\ast })$. 
The universal energy functional at the equilibrium densities is given by
\begin{eqnarray}
\label{eq37}
F\left[ {n_{0} ,{\rm {\bf j}}_{p0}^{(T)} ,\Delta_{0} ,\Delta_{0}^{\ast}} \right]
&=&
\mathop{\mbox{Min}}\limits_{\hat{{\rho }}\to n_{0} ,{\rm {\bf 
j}}_{p0}^{(T)} ,\Delta_{0} ,\Delta_{0}^{\ast }} \mbox{Tr}\left\{ 
{\hat{{\rho }}\left( {\hat{{T}}+\hat{{W}}_{1} +\hat{{W}}_{2} } 
\right)+{\frac{1}{\beta }}\hat{{\rho }}\ln \hat{{\rho }}} \right\} \nonumber \\ 
&=&
\mbox{Tr}\left\{ {\hat{{\rho }}_{\min } \left[ {n_{0} ,{\rm {\bf 
j}}_{p0}^{(T)} ,\Delta_{0} ,\Delta_{0}^{\ast }} \right]\left( 
{\hat{{T}}+\hat{{W}}_{1} +\hat{{W}}_{2} } \right)} \right. \nonumber \\
&&\left. {+{\frac{1}{\beta }}\hat{{\rho }}_{\min } \left[ 
{n_{0} ,{\rm {\bf j}}_{p0}^{(T)} ,\Delta_{0} ,\Delta_{0}^{\ast }} 
\right]\ln \hat{{\rho }}_{\min } \left[ {n_{0} ,{\rm {\bf j}}_{p0}^{(T)} 
,\Delta_{0} ,\Delta_{0}^{\ast }} \right]} \right\} , 
\end{eqnarray}
where the minimizing statistical operator is referred to as $\hat{{\rho 
}}_{\min } \left[ {n_{0} ,{\rm {\bf j}}_{p0}^{(T)} ,\Delta_{0} ,\Delta_{0} 
^{\ast }} \right]$. Taking into account Gibbs's variational principle, i.e., 
Eq. (\ref{eq24}), the following relation holds:
\begin{eqnarray}
\label{eq38}
&\mbox{Tr}&\left\{ {\hat{{\rho }}_{0} \left( {\hat{{T}}+\hat{{W}}_{1} 
+\hat{{W}}_{2} +\hat{{V}}_{1} +\hat{{V}}_{2} +\hat{{V}}_{3} -\mu \hat{{N}}} 
\right)+{\frac{1}{\beta }}\hat{{\rho }}_{0} \ln \hat{{\rho }}_{0} } \right\} 
\nonumber \\ 
 \le &\mbox{Tr}&\left\{ {\hat{{\rho }}_{\min } \left[ {n_{0} ,{\rm {\bf 
j}}_{p0}^{(T)} ,\Delta_{0} ,\Delta_{0}^{\ast }} \right]\left( 
{\hat{{T}}+\hat{{W}}_{1} +\hat{{W}}_{2} +\hat{{V}}_{1} +\hat{{V}}_{2} 
+\hat{{V}}_{3} -\mu \hat{{N}}} \right)} \right. \nonumber \\ 
&&\left. {+{\frac{1}{\beta }}\hat{{\rho }}_{\min } \left[ 
{n_{0} ,{\rm {\bf j}}_{p0}^{(T)} ,\Delta_{0} ,\Delta_{0}^{\ast }} 
\right]\ln \hat{{\rho }}_{\min } \left[ {n_{0} ,{\rm {\bf j}}_{p0}^{(T)} 
,\Delta_{0} ,\Delta_{0}^{\ast }} \right]} \right\} . 
\end{eqnarray}
Since both $\hat{{\rho }}_{0} $ and $\hat{{\rho }}_{\min } \left[ {n_{0} 
,{\rm {\bf j}}_{p0}^{(T)} ,\Delta_{0} ,\Delta_{0}^{\ast }} \right]$ yield 
the correct densities $n_{0} ,\,\,{\rm {\bf j}}_{p0}^{(T)} ,\,\,\Delta_{0} 
,\,\,\Delta_{0}^{\ast }$, Eq. (\ref{eq38}) becomes to
\begin{equation}
\label{eq39}
\mbox{Tr}\left\{ {\hat{{\rho }}_{0} \left( {\hat{{T}}+\hat{{W}}_{1} 
+\hat{{W}}_{2} } \right)+{\frac{1}{\beta }}\hat{{\rho }}_{0} \ln \hat{{\rho 
}}_{0} } \right\}\le F\left[ {n_{0} ,{\rm {\bf j}}_{p0}^{(T)} ,\Delta_{0} 
,\Delta_{0}^{\ast }} \right].
\end{equation}
From the definition of $F\left[ {n_{0} ,{\rm {\bf j}}_{p0}^{(T)} ,\Delta 
_{0} ,\Delta_{0}^{\ast }} \right]$, i.e., Eq. (\ref{eq37}), only an equal sign is 
satisfied in Eq. (\ref{eq39}). Namely we have
\begin{eqnarray}
\label{eq40}
\!\!\!\!\!\!\!\!\!\!\!\!\mbox{Tr}\left\{ {\hat{{\rho }}_{0} \left( {\hat{{T}}+\hat{{W}}_{1} 
+\hat{{W}}_{2} } \right)+{\frac{1}{\beta }}\hat{{\rho }}_{0} \ln \hat{{\rho 
}}_{0} } \right\}
&=&
\mbox{Tr}\left\{ {\hat{{\rho }}_{\min } \left[ {n_{0} 
,{\rm {\bf j}}_{p0}^{(T)} ,\Delta_{0} ,\Delta_{0}^{\ast }} \right]\left( 
{\hat{{T}}+\hat{{W}}_{1} +\hat{{W}}_{2} } \right)} \right. \nonumber \\ 
&&\!\!\!\!\!\!\!\!\!\!\!\!\!\!\!\!\
\left. {+{\frac{1}{\beta }}\hat{{\rho }}_{\min } \left[ 
{n_{0} ,{\rm {\bf j}}_{p0}^{(T)} ,\Delta_{0} ,\Delta_{0}^{\ast }} 
\right]\ln \hat{{\rho }}_{\min } \left[ {n_{0} ,{\rm {\bf j}}_{p0}^{(T)} 
,\Delta_{0} ,\Delta_{0}^{\ast }} \right]} \right\} . 
\end{eqnarray}
Using the fact that both $\hat{{\rho }}_{0} $ and $\hat{{\rho }}_{\min } 
\left[ {n_{0} ,{\rm {\bf j}}_{p0}^{(T)} ,\Delta_{0} ,\Delta_{0}^{\ast }} 
\right]$ yield the same densities $(n_{0} ,\,\,{\rm {\bf j}}_{p0}^{(T)} 
,\,\,\Delta_{0} ,\,\,\Delta_{0}^{\ast })$, and using Eq. (\ref{eq40}), only an 
equal sign holds also in Eq. (\ref{eq38}):
\begin{eqnarray}
\label{eq41}
&&\mbox{Tr}\left\{ {\hat{{\rho }}_{0} \left( {\hat{{T}}+\hat{{W}}_{1} 
+\hat{{W}}_{2} +\hat{{V}}_{1} +\hat{{V}}_{2} +\hat{{V}}_{3} -\mu \hat{{N}}} 
\right)+{\frac{1}{\beta }}\hat{{\rho }}_{0} \ln \hat{{\rho }}_{0} } \right\} 
\nonumber \\ 
&=&
\mbox{Tr}\left\{ {\hat{{\rho }}_{\min } \left[ {n_{0} ,{\rm {\bf 
j}}_{p0}^{(T)} ,\Delta_{0} ,\Delta_{0}^{\ast }} \right]\left( 
{\hat{{T}}+\hat{{W}}_{1} +\hat{{W}}_{2} +\hat{{V}}_{1} +\hat{{V}}_{2} 
+\hat{{V}}_{3} -\mu \hat{{N}}} \right)} \right. \nonumber \\ 
&&\left. {+{\frac{1}{\beta }}\hat{{\rho }}_{\min } \left[ 
{n_{0} ,{\rm {\bf j}}_{p0}^{(T)} ,\Delta_{0} ,\Delta_{0}^{\ast }} 
\right]\ln \hat{{\rho }}_{\min } \left[ {n_{0} ,{\rm {\bf j}}_{p0}^{(T)} 
,\Delta_{0} ,\Delta_{0}^{\ast }} \right]} \right\} . 
\end{eqnarray}
The LHS of Eq. (\ref{eq41}) is exactly the correct grand potential $J_{0} $. 
Considering Gibbs's variational theorem Eq. (\ref{eq24}), we finally obtain
\begin{equation}
\label{eq42}
\hat{{\rho }}_{\min } \left[ {n_{0} ,{\rm {\bf j}}_{p0}^{(T)} ,\Delta_{0} 
,\Delta_{0}^{\ast }} \right]=\hat{{\rho }}_{0} .
\end{equation}
It follows that the correct density matrix $\hat{{\rho }}_{0} 
\,\,\,(=\hat{{\rho }}_{\min } \left[ {n_{0} ,{\rm {\bf j}}_{p0}^{(T)} 
,\Delta_{0} ,\Delta_{0}^{\ast }} \right])$ is uniquely determined by the 
correct densities $(n_{0} ,\,\,{\rm {\bf j}}_{p0}^{(T)} ,\,\,\Delta_{0} 
,\,\,\Delta_{0}^{\ast })$ via Eq. (\ref{eq37}), and vice versa due to Eqs. (\ref{eq33}) -- 
(\ref{eq36}). Thus, it is successfully proved that the one-to-one correspondence 
between the statistical operator $\hat{{\rho }}_{0} $ and densities $(n_{0} 
,\,\,{\rm {\bf j}}_{p0}^{(T)} ,\,\,\Delta_{0} ,\,\,\Delta_{0}^{\ast })$ holds.
%
%********************* Sec. IV ******************************************
\section{Single-particle equation in the reference system}
\label{secIV}
In order to reproduce the equilibrium densities $(n_{0} ,\,\,{\rm {\bf 
j}}_{p0}^{(T)} ,\,\,\Delta_{0} ,\,\,\Delta_{0}^{\ast })$, we shall 
introduce the reference system in a similar way to the ECS theory \cite{39,40}. 
The reference system is the noninteracting system in which there exist the 
mean-field potentials applied so as to make basic variables coincide with 
the equilibrium densities. 

Since it is in the quadratic form in terms of the field operator of 
electrons, the Hamiltonian of the reference system can be diagonalized in 
terms of the noninteracting fermion quasiparticle via the so-called 
Bogoliubov-Valatin (BV) transformation \cite{56,57}. The matrix elements of the 
BV transformation can be obtained by solving the single-particle equation 
which is called the BdG-KS equation. In this section we also derive the 
BdG-KS equation and confirm its solutions to be satisfied with orthonormal 
and complete properties caused from the property of the BV transformation. 
%
%******************* subsec.IV-A *********************************************
\subsection{Reference system}
\label{secIV-A}
We introduce the noninteracting system as the reference system in which the 
effective mean-field potentials are applied instead of the interaction 
energy term consisting of four field operators of electrons. The Hamiltonian 
of the reference system is given by
\begin{eqnarray}
\label{eq43}
\hat{{H}}_{s} 
&=&
\int {\psi^{\dag }({\rm {\bf r}}\zeta )\left[ 
{{\frac{1}{2m}}\left\{ {{\rm {\bf p}}+e{\rm {\bf A}}_{s} ({\rm {\bf r}})} 
\right\}^{2}} \right]} \psi ({\rm {\bf r}}\zeta )d^{3}rd\zeta \nonumber \\ 
&&+
\int {v_{s} ({\rm {\bf r}})\psi^{\dag }({\rm {\bf r}}\zeta )\psi ({\rm 
{\bf r}}\zeta )} d^{3}rd\zeta \nonumber \\ 
&&+
\int\!\!\!\int {D_{s}^{\ast } ({\rm {\bf r}}\zeta, {\rm {\bf {r}'}}{\zeta 
}')\hat{{\Delta }}({\rm {\bf r}}\zeta, {\rm {\bf {r}'}}{\zeta }')d^{3}rd\zeta 
d^{3}{r}'d{\zeta }'} \nonumber \\ 
&&+
\int\!\!\!\int {D_{s} ({\rm {\bf r}}\zeta, {\rm {\bf {r}'}}{\zeta 
}')\hat{{\Delta }}^{\dag }({\rm {\bf r}}\zeta, {\rm {\bf {r}'}}{\zeta 
}')d^{3}rd\zeta d^{3}{r}'d{\zeta }'} , 
\end{eqnarray}
where $v_{s} ({\rm {\bf r}})$, ${\rm {\bf A}}_{s} ({\rm {\bf r}})$, 
$D_{s} ({\rm {\bf r}}\zeta, {\rm {\bf {r}'}}{\zeta }')$ and $D_{s}^{\ast } 
({\rm {\bf r}}\zeta, {\rm {\bf {r}'}}{\zeta }')$ are effective mean-field 
potentials. Especially $D_{s} ({\rm {\bf r}}\zeta, {\rm {\bf {r}'}}{\zeta 
}')$ is called the effective pair potential which induces the OPSS to be 
nonzero when the system is in the superconducting state. The concrete forms 
of these effective potentials are determined by using the HK theorem, which 
will be shown in Sec. V.

The essential point about the reference system is that the Hamiltonian (\ref{eq43}) 
includes the pair potential terms, i.e., the 3rd and 4th terms, which 
explicitly breaks the conservation of the electron number. In other words, 
we intentionally prepare for the reference system which can express the 
symmetry breaking state of the $U$(\ref{eq1}) gauge. This device for the reference 
system is similar to that for the mean-field approximation of the BCS theory 
\cite{53}.

Let us consider the transformation from the system of electrons to that of 
the fermion quasiparticles, annihilation and creation operators of which are 
denoted as $\gamma_{i} $ and $\gamma_{i}^{\dag } $, respectively. Since it 
is quadratic in terms of the field operator of electrons, the Hamiltonian 
(\ref{eq43}) can be diagonalized in terms of the fermion quasiparticle via the BV 
transformation \cite{56,57}. Suppose that the transformation is given by 
\begin{equation}
\label{eq44}
\begin{array}{l}
 \psi ({\rm {\bf r}}\zeta )=\sum\limits_i {u_{i} ({\rm {\bf r}}\zeta 
)\gamma_{i} } +\sum\limits_j {v_{j} ({\rm {\bf r}}\zeta )\gamma_{j}^{\dag 
} } , \\ 
 \psi^{\dag }({\rm {\bf r}}\zeta )=\sum\limits_i {u_{i}^{\ast } ({\rm {\bf 
r}}\zeta )\gamma_{i}^{\dag } } +\sum\limits_j {v_{j}^{\ast } ({\rm {\bf 
r}}\zeta )\gamma_{j} } , \\ 
 \end{array}
\end{equation}
where $u_{i} ({\rm {\bf r}}\zeta )$ and $v_{i} ({\rm {\bf r}}\zeta )$ 
correspond to the matrix elements of the BV transformation. They are 
determined by requiring $\hat{{H}}_{s} $ to be diagonalized in terms of 
$\gamma_{i} $ and $\gamma_{i}^{\dag } $. 

Substituting Eq. (\ref{eq44}) into the aniticommutation relations for $\psi ({\rm 
{\bf r}}\zeta )$ and $\psi^{\dag }({\rm {\bf r}}\zeta )$, and using the 
anticommutation relations for $\gamma_{i} $ and $\gamma_{i}^{\dag } $, the 
following relations hold:
\begin{eqnarray}
\label{eq45}
&&\sum\limits_i {\left\{ {u_{i}^{\ast } ({\rm {\bf r}}\zeta )u_{i} ({\rm {\bf 
{r}'}}{\zeta }')+v_{i}^{\ast } ({\rm {\bf r}}\zeta )v_{i} ({\rm {\bf 
{r}'}}{\zeta }')} \right\}} =\delta ({\rm {\bf r}}-{\rm {\bf {r}'}})\delta 
_{\zeta {\zeta }'} , \\
\label{eq46}
&&\sum\limits_i {\left\{ {u_{i} ({\rm {\bf r}}\zeta )v_{i} ({\rm {\bf 
{r}'}}{\zeta }')+v_{i} ({\rm {\bf r}}\zeta )u_{i} ({\rm {\bf {r}'}}{\zeta 
}')} \right\}} =0.
\end{eqnarray}
In other words, Eqs. (\ref{eq45}) and (\ref{eq46}) have to be satisfied so that Eq. (\ref{eq44}) is 
consistent with both anticommutation relations for $\psi ({\rm {\bf r}}\zeta 
)$ and $\psi^{\dag }({\rm {\bf r}}\zeta )$ and those for $\gamma_{i} $ and 
$\gamma_{i}^{\dag } $. For the purpose of reference, the property of the BV 
transformation Eq. (\ref{eq44}) is given in the Appendix. 

%****************** subsection IV-B **************************************
\subsection{BdG-KS equation}
\label{secIV-B}
In this subsection, we derive the conditions on $u_{i} ({\rm {\bf 
r}}\zeta )$ and $v_{i} ({\rm {\bf r}}\zeta )$ for diagonalyzing the 
Hamiltonian $\hat{{H}}_{s} $ in terms of $\gamma_{i} $ and $\gamma 
_{i}^{\dag } $. Suppose that the diagonalized Hamiltonian is given by
\begin{equation}
\label{eq47}
\hat{{H}}_{s} -\mu \hat{{N}}=E_{g} +\sum\limits_i {E_{i} \gamma_{i}^{\dag } 
\gamma_{i} } ,
\end{equation}
where $E_{g} $ and $E_{i} $ are the ground-state energy and excited-state 
energies, respectively. Using Eq. (\ref{eq47}) and anticommutation relations for 
$\gamma_{i} $ and $\gamma_{i}^{\dag } $, we have
\begin{eqnarray}
\label{eq48}
&&\left[ {\,\gamma_{i} ,\,\hat{{H}}_{s} -\mu \hat{{N}}} \right]
=E_{i} \gamma _{i} , \\
\label{eq49}
&&\left[ {\gamma_{i}^{\dag } ,\,\,\hat{{H}}_{s} -\mu \hat{{N}}} 
\right]=-E_{i} \gamma_{i}^{\dag } .
\end{eqnarray}
Also, using Eq. (\ref{eq43}) and aniticommutation relations for 
$\psi ({\rm {\bf r}}\zeta )$ and 
$\psi^{\dag }({\rm {\bf r}}\zeta )$, we have
\begin{eqnarray}
\label{eq50}
\left[ {\psi ({\rm {\bf r}}\zeta ),\,\,\hat{{H}}_{s} -\mu \hat{{N}}} 
\right]
&=&
\left( {h_{s}^{{\rm {\bf r}}} -\mu } \right)\psi ({\rm {\bf r}}\zeta 
) \nonumber \\ 
&&+\int {\left\{ {D_{s} ({\rm {\bf r}}\zeta ,{\rm {\bf {r}'}}{\zeta }')-D_{s} 
({\rm {\bf {r}'}}{\zeta }',{\rm {\bf r}}\zeta )} \right\}\psi^{\dag }({\rm 
{\bf {r}'}}{\zeta }')d^{3}{r}'d{\zeta }'} , \\
\label{eq51}
\left[ {\psi^{\dag }({\rm {\bf r}}\zeta ),\,\,\hat{{H}}_{s} -\mu 
\hat{{N}}} \right]
&=&
-\int {\psi^{\dag }({\rm {\bf {r}'}}{\zeta }')\left( 
{h_{s}^{{\rm {\bf {r}'}}} -\mu } \right)\delta ({\rm {\bf r}}-{\rm {\bf 
{r}'}})\delta_{\zeta {\zeta }'} d^{3}{r}'d{\zeta }'} \nonumber \\ 
&&+\int {\left\{ {D_{s}^{\ast } ({\rm {\bf {r}'}}{\zeta }',{\rm {\bf r}}\zeta 
)-D_{s}^{\ast } ({\rm {\bf r}}\zeta ,{\rm {\bf {r}'}}{\zeta }')} 
\right\}\psi ({\rm {\bf {r}'}}{\zeta }')d^{3}{r}'d{\zeta }'} , 
\end{eqnarray}
where $h_{s}^{{\rm {\bf r}}} $ is defined as
\begin{equation}
\label{eq52}
h_{s}^{{\rm {\bf r}}} ={\frac{1}{2m}}\left\{ {{\rm {\bf p}}+e{\rm {\bf 
A}}_{s} ({\rm {\bf r}})} \right\}^{2}+v_{s} ({\rm {\bf r}}).
\end{equation}
Substituting Eq. (\ref{eq44}) into Eq. (\ref{eq50}), and using Eqs. (\ref{eq48}) and (\ref{eq49}), we 
finally obtain the single-particle equation with which $u_{i} ({\rm {\bf 
r}}\zeta )$ and $v_{i} ({\rm {\bf r}}\zeta )$ should be satisfied:
\begin{equation}
\label{eq53}
\begin{array}{l}
 \left( {h_{s}^{{\rm {\bf r}}} -\mu } \right)u_{i} ({\rm {\bf r}}\zeta 
)+\int {\tilde{{D}}_{s} ({\rm {\bf r}}\zeta ,{\rm {\bf {r}'}}{\zeta 
}')v_{i}^{\ast } ({\rm {\bf {r}'}}{\zeta }')d^{3}{r}'d{\zeta }'} =E_{i} 
u_{i} ({\rm {\bf r}}\zeta ), \\ 
 -\left( {h_{s}^{{\rm {\bf r}}} -\mu } \right)v_{i} ({\rm {\bf r}}\zeta 
)-\int {\tilde{{D}}_{s} ({\rm {\bf r}}\zeta ,{\rm {\bf {r}'}}{\zeta 
}')u_{i}^{\ast } ({\rm {\bf {r}'}}{\zeta }')d^{3}{r}'d{\zeta }'} =E_{i} 
v_{i} ({\rm {\bf r}}\zeta ), \\ 
 \end{array}
\end{equation}
where
\begin{equation}
\label{eq54}
\tilde{{D}}_{s} ({\rm {\bf r}}\zeta ,{\rm {\bf {r}'}}{\zeta }')=D_{s} ({\rm 
{\bf r}}\zeta ,{\rm {\bf {r}'}}{\zeta }')-D_{s} ({\rm {\bf {r}'}}{\zeta 
}',{\rm {\bf r}}\zeta ).
\end{equation}
The solutions of Eq. (\ref{eq53}) correspond to the matrix elements of the BV 
transformation which makes the Hamiltonian Eq. (\ref{eq43}) be diagonalized in the 
form of Eq. (\ref{eq47}). 
The equation (\ref{eq53}) is called the BdG-KS equation \cite{44,48}. 
As shown in the next subsection, the eigenvalue $E_{i} $ is a real number. 
Using this result, it is shown that Eq. (\ref{eq53}) is also obtained from starting 
with Eq. (\ref{eq51}) instead of Eq. (\ref{eq50}).

%
%****************** subsection IV-C **************************************
\subsection{Orthonormality for the solution of the BdG-KS equation}
\label{secIV-C}
In this subsection, we discuss the orthonormal properties of the solution of 
the BdG-KS equation that are used in developing the approximate form of the 
xc energy functional (Sec. VI). First we show that the eigenvalue $E_{i} $ 
of Eq. (\ref{eq53}) is a real number. Multiplying both sides of the first (second) 
equation of Eq. (\ref{eq53}) by $u_{i}^{\ast } ({\rm {\bf r}}\zeta )$ ($v_{i}^{\ast 
} ({\rm {\bf r}}\zeta ))$ from the left, and integrating with respect to 
${\rm {\bf r}}$ and $\zeta $, yield
\begin{eqnarray}
\label{eq55}
&&\int {u_{i}^{\ast } ({\rm {\bf r}}\zeta )\left( {h_{s}^{{\rm {\bf r}}} -\mu 
} \right)u_{i} ({\rm {\bf r}}\zeta )d^{3}rd\zeta } \nonumber \\ 
&&+\int\!\!\!\int {u_{i}^{\ast } ({\rm {\bf r}}\zeta )\tilde{{D}}_{s} ({\rm 
{\bf r}}\zeta ,{\rm {\bf {r}'}}{\zeta }')v_{i}^{\ast } ({\rm {\bf 
{r}'}}{\zeta }')d^{3}{r}'d{\zeta }'d^{3}rd\zeta } =E_{i} \int {\left| {u_{i} 
({\rm {\bf r}}\zeta )} \right|^{2}d^{3}rd\zeta } , 
\end{eqnarray}
\begin{eqnarray}
\label{eq56}
&&-\int {v_{i}^{\ast } ({\rm {\bf r}}\zeta )\left( {h_{s}^{{\rm {\bf r}}} 
-\mu } \right)v_{i} ({\rm {\bf r}}\zeta )d^{3}rd\zeta } \nonumber \\ 
&&+\int\!\!\!\int {u_{i}^{\ast } ({\rm {\bf r}}\zeta )\tilde{{D}}_{s} ({\rm 
{\bf r}}\zeta ,{\rm {\bf {r}'}}{\zeta }')v_{i}^{\ast } ({\rm {\bf 
{r}'}}{\zeta }')d^{3}{r}'d{\zeta }'d^{3}rd\zeta } =E_{i} \int {\left| {v_{i} 
({\rm {\bf r}}\zeta )} \right|^{2}d^{3}rd\zeta } , 
\end{eqnarray}
respectively. Subtracting Eq. (\ref{eq56}) from Eq. (\ref{eq55}) on both sides, we have
\begin{eqnarray}
\label{eq57}
&&\int {u_{i}^{\ast } ({\rm {\bf r}}\zeta )\left( {h_{s}^{{\rm {\bf r}}} -\mu 
} \right)u_{i} ({\rm {\bf r}}\zeta )d^{3}rd\zeta } +\int {v_{i}^{\ast } 
({\rm {\bf r}}\zeta )\left( {h_{s}^{{\rm {\bf r}}} -\mu } \right)v_{i} ({\rm 
{\bf r}}\zeta )d^{3}rd\zeta } \nonumber \\ 
&=&
E_{i} \int {\left\{ {\left| {u_{i} ({\rm {\bf r}}\zeta )} 
\right|^{2}-\left| {v_{i} ({\rm {\bf r}}\zeta )} \right|^{2}} 
\right\}d^{3}rd\zeta } . 
\end{eqnarray}
Since both terms of the LHS are real numbers due to the Hermitian property 
of Eq. (\ref{eq52}) and since the integration of the RHS is a real number, $E_{i} $ 
is necessarily a real number. Using this result, Eq. (\ref{eq53}) is rewritten as
\begin{equation}
\label{eq58}
\hat{{\Lambda }}\left( {{\begin{array}{*{20}c}
 {u_{i} ({\rm {\bf r}}\zeta )} \hfill \\
 {v_{i}^{\ast } ({\rm {\bf r}}\zeta )} \hfill \\
\end{array} }} \right)=E_{i} \left( {{\begin{array}{*{20}c}
 {u_{i} ({\rm {\bf r}}\zeta )} \hfill \\
 {v_{i}^{\ast } ({\rm {\bf r}}\zeta )} \hfill \\
\end{array} }} \right),
\end{equation}
where $\hat{{\Lambda }}$ is a $2\times 2$ matrix defined as
\begin{equation}
\label{eq59}
\hat{{\Lambda }}=\left( {{\begin{array}{*{20}c}
 {h_{s}^{{\rm {\bf r}}} -\mu } \hfill & {\hat{{D}}_{s} } \hfill \\
 {-\hat{{D}}_{s}^{\ast } } \hfill & {-\left( {h_{s}^{{\rm {\bf r}}} -\mu } 
\right)^{\ast }} \hfill \\
\end{array} }} \right),
\end{equation}
and where $\hat{{D}}_{s} $ is defined as the operator which acts the 
function $f({\rm {\bf r}}\zeta )$ such that
\begin{equation}
\label{eq60}
\hat{{D}}_{s} f({\rm {\bf r}}\zeta )=\int {\tilde{{D}}_{s} ({\rm {\bf 
r}}\zeta ,{\rm {\bf {r}'}}{\zeta }')f({\rm {\bf {r}'}}{\zeta 
}')d^{3}{r}'d{\zeta }'} .
\end{equation}
The solution 
$\left( {{\begin{array}{*{20}c}
 {u_{i} ({\rm {\bf r}}\zeta )} \hfill \\
 {v_{i}^{\ast } ({\rm {\bf r}}\zeta )} \hfill \\
\end{array} }} \right)$ 
can be regarded as the eigenstate of Eq. (\ref{eq58}). 
It is easily shown that the matrix elements
\begin{equation}
\label{eq61}
\Lambda_{ij} =\int {\left( {u_{i} ({\rm {\bf r}}\zeta )\,\,\,v_{i}^{\ast } 
({\rm {\bf r}}\zeta )} \right)^{\ast }\hat{{\Lambda }}\left( 
{{\begin{array}{*{20}c}
 {u_{j} ({\rm {\bf r}}\zeta )} \hfill \\
 {v_{j}^{\ast } ({\rm {\bf r}}\zeta )} \hfill \\
\end{array} }} \right)d^{3}rd\zeta } ,
\end{equation}
are satisfied with the Hermitian property $\Lambda_{ij} =\Lambda 
_{ji}^{\ast } $. Using this property, it is also shown in a usual way \cite{58} 
that the eigenstates of Eq. (\ref{eq58}) can be chosen to have the orthonormal 
property
\begin{equation}
\label{eq62}
\int {\left\{ {u_{i}^{\ast } ({\rm {\bf r}}\zeta )u_{j} ({\rm {\bf r}}\zeta 
)+v_{i} ({\rm {\bf r}}\zeta )v_{j}^{\ast } ({\rm {\bf r}}\zeta )} 
\right\}d^{3}rd\zeta } =\delta_{ij} .
\end{equation}

Further, the other type of orthonormal property is obtained by utilizing the 
solution of the BdG-KS equation. It is confirmed that the BdG-KS equation 
(\ref{eq58}) can be rewritten as
\begin{equation}
\label{eq63}
\hat{{\Lambda }}\left( {{\begin{array}{*{20}c}
 {v_{i} ({\rm {\bf r}}\zeta )} \hfill \\
 {u_{i}^{\ast } ({\rm {\bf r}}\zeta )} \hfill \\
\end{array} }} \right)=-E_{i} \left( {{\begin{array}{*{20}c}
 {v_{i} ({\rm {\bf r}}\zeta )} \hfill \\
 {u_{i}^{\ast } ({\rm {\bf r}}\zeta )} \hfill \\
\end{array} }} \right).
\end{equation}
Namely, 
$\left( {{\begin{array}{*{20}c}
 {v_{i} ({\rm {\bf r}}\zeta )} \hfill \\
 {u_{i}^{\ast } ({\rm {\bf r}}\zeta )} \hfill \\
\end{array} }} \right)$ 
is also the solution of the BdG-KS equation, and the 
corresponding eigenvalue is $-E_{i} $. Since the excitation energy of the 
quasiparticle is supposed to be positive, i.e., $E_{i} \ne -E_{j} $, we have 
\begin{equation}
\label{eq64}
\int {\left\{ {u_{i}^{\ast } ({\rm {\bf r}}\zeta )v_{j} ({\rm {\bf r}}\zeta 
)+v_{i} ({\rm {\bf r}}\zeta )u_{j}^{\ast } ({\rm {\bf r}}\zeta )} 
\right\}d^{3}rd\zeta } =0.
\end{equation}
In the special case $i=j$, Eq. (\ref{eq64}) becomes to
\begin{equation}
\label{eq65}
\int {u_{i}^{\ast } ({\rm {\bf r}}\zeta )v_{i} ({\rm {\bf r}}\zeta 
)d^{3}rd\zeta } =0.
\end{equation}

Equations (\ref{eq45}) and (\ref{eq46}) can be regarded as a kind of completeness of the 
solutions of Eq. (\ref{eq58}) because such relations contain the summation on the 
number $i$ specifying the eigenstates of Eq. (\ref{eq58}). On the other hand, Eqs. 
(\ref{eq62}), (\ref{eq64}) and (\ref{eq65}) can be regarded as the orthonormal property of the 
solutions of Eq. (\ref{eq58}).
%

%********************* Sec. V ******************************************
\section{Effective mean-field potentials}
\label{secV}
In this section we present the explicit forms of the effective mean-field 
potentials $v_{s} ({\rm {\bf r}})$, ${\rm {\bf A}}_{s} ({\rm {\bf r}})$, 
$D_{s} ({\rm {\bf r}}\zeta, {\rm {\bf {r}'}}{\zeta }')$ and $D_{s}^{\ast } 
({\rm {\bf r}}\zeta, {\rm {\bf {r}'}}{\zeta }')$. They are determined by 
requiring the basic variables calculated in the reference system to be in 
accordance with the equilibrium densities.

%********************* Sec. V A ******************************************
\subsection{The HK theorem for the reference system}
\label{secV-A}
First, we consider the HK theorem for the reference system. Since the proof 
is similar to that of the real system (Sec. III), we shall show only the 
outline of this theorem. The Hamiltonian of the reference system is given by 
Eq. (\ref{eq43}). In a similar way to Eq. (\ref{eq5}), neglecting the surface integral at 
the infinite distance and adopting the Coulomb gauge given by $\nabla \cdot 
{\rm {\bf A}}_{s} ({\rm {\bf r}})=0$, Eq. (\ref{eq43}) is rewritten as 
\begin{equation}
\label{eq66}
\hat{{H}}_{s} =\hat{{T}}+\hat{{V}}_{1s} +\hat{{V}}_{2s} +\hat{{V}}_{3s} 
+\hat{{V}}_{Ds} ,
\end{equation}
where $\hat{{T}}$ is given by Eq. (\ref{eq6}), and where 
\begin{eqnarray}
\label{eq67}
&&\hat{{V}}_{1s} =\int {v_{s} ({\rm {\bf r}})\hat{{n}}({\rm {\bf r}})} d^{3}r,
\\
\label{eq68}
&&\hat{{V}}_{2s} =e\int {{\rm {\bf A}}_{s} ({\rm {\bf r}})\cdot {\rm {\bf 
\hat{{j}}}}_{p}^{(T)} ({\rm {\bf r}})d^{3}r} ,
\\
\label{eq69}
&&\hat{{V}}_{3s} ={\frac{e^{2}}{2m}}\int {{\rm {\bf A}}_{s} ({\rm {\bf 
r}})^{2}\hat{{n}}({\rm {\bf r}})d^{3}r} ,
\\
\label{eq70}
&&\hat{{V}}_{Ds} =\int\!\!\!\int {D_{s}^{\ast } ({\rm {\bf r}}\zeta, {\rm {\bf 
{r}'}}{\zeta }')\hat{{\Delta }}({\rm {\bf r}}\zeta, {\rm {\bf {r}'}}{\zeta 
}')d^{3}rd\zeta d^{3}{r}'d{\zeta }'} \nonumber \\
&&\ \ \ \ \ \ \ \ +\int\!\!\!\int {D_{s} ({\rm {\bf 
r}}\zeta, {\rm {\bf {r}'}}{\zeta }')\hat{{\Delta }}^{\ast }({\rm {\bf 
r}}\zeta, {\rm {\bf {r}'}}{\zeta }')d^{3}rd\zeta d^{3}{r}'d{\zeta }'} .
\end{eqnarray}
The universal energy functional for the reference system is defined as
\begin{equation}
\label{eq71}
F_{s} \left[ {n,{\rm {\bf j}}_{p}^{(T)} ,\Delta ,\Delta^{\ast }} 
\right]=\mathop{\mbox{Min}}\limits_{\hat{{\rho }}\to n,{\rm {\bf j}}_{p}^{(T)} 
,\Delta ,\Delta^{\ast }} \mbox{Tr}\left\{ {\hat{{\rho 
}}\,\hat{{T}}+{\frac{1}{\beta }}\hat{{\rho }}\ln \hat{{\rho }}} \right\}.
\end{equation}
Gibbs's variational principle holds also for the reference system. If we 
define the following functional
%
%\[
\begin{equation}
J_{s} \left[ {\hat{{\rho }}} \right]=\mbox{Tr}\left\{ {\hat{{\rho }}\left( 
{\hat{{H}}_{s} -\mu \hat{{N}}} \right)+{\frac{1}{\beta }}\hat{{\rho }}\ln 
\hat{{\rho }}} \right\},
\end{equation}
%\]
%
then the minimum point $J_{s0} $ exists at the statistical operator 
$\hat{{\rho }}_{s0} $:
\begin{equation}
\label{eq72}
J_{s0} =\mathop{\mbox{Min}}\limits_{\hat{{\rho }}} J_{s} \left[ {\hat{{\rho }}} 
\right]=J_{s} \left[ {\hat{{\rho }}_{s0} } \right],
\end{equation}
where
\begin{equation}
\label{eq73}
\hat{{\rho }}_{s0} ={\frac{e^{-\beta \left( {\hat{{H}}_{s} -\mu \hat{{N}}} 
\right)}}{\Xi_{s} }}
\end{equation}
with $\Xi_{s} =\mbox{Tr}\left( {e^{-\beta \left( {\hat{{H}}_{s} -\mu 
\hat{{N}}} \right)}} \right)$. In a similar way to the case of the real 
system, the variational principles Eq. (\ref{eq72}) can be rewritten as the 
variational principle with respect to the densities chosen as the basic 
variables, i.e., ($n,{\rm {\bf j}}_{p}^{(T)} ,\Delta ,\Delta^{\ast })$. We have
\begin{equation}
\label{eq74}
J_{s0} =\mathop{\mbox{Min}}\limits_{n,{\rm {\bf j}}_{p}^{(T)} ,\Delta ,\Delta 
^{\ast }} J_{s}^{v_{s} -\mu ,\,{\rm {\bf A}}_{s} ,D_{s} ,D_{s}^{\ast } } 
\left[ {n,{\rm {\bf j}}_{p}^{(T)} ,\Delta ,\Delta^{\ast }} \right],
\end{equation}
where 
\begin{eqnarray}
\label{eq75}
J_{s}^{v_{s} -\mu ,\,{\rm {\bf A}}_{s} ,D_{s} ,D_{s}^{\ast } } \left[ 
{n,{\rm {\bf j}}_{p}^{(T)} ,\Delta ,\Delta^{\ast }} \right]
&=&
F_{s} \left[ {n,{\rm {\bf j}}_{p}^{(T)} ,\Delta ,\Delta^{\ast }} \right]+\int {\left\{ 
{v_{s} ({\rm {\bf r}})-\mu } \right\}n({\rm {\bf r}})} d^{3}r \nonumber \\ 
&+&
e\int {{\rm {\bf A}}_{s} ({\rm {\bf r}})\cdot {\rm {\bf j}}_{p}^{(T)} 
({\rm {\bf r}})d^{3}r+} {\frac{e^{2}}{2m}}\int {{\rm {\bf A}}_{s} ({\rm {\bf 
r}})^{2}n({\rm {\bf r}})d^{3}r} \nonumber \\ 
&+&
\int\!\!\!\int {D_{s}^{\ast } ({\rm {\bf r}}\zeta, {\rm {\bf {r}'}}{\zeta 
}')\Delta ({\rm {\bf r}}\zeta, {\rm {\bf {r}'}}{\zeta }')d^{3}rd\zeta 
d^{3}{r}'d{\zeta }'} \nonumber \\ 
&+&
\int\!\!\!\int {D_{s} ({\rm {\bf r}}\zeta, {\rm {\bf {r}'}}{\zeta 
}')\Delta^{\ast }({\rm {\bf r}}\zeta, {\rm {\bf {r}'}}{\zeta }')d^{3}rd\zeta 
d^{3}{r}'d{\zeta }'} . 
\end{eqnarray}
The minimizing densities searched in Eq. (\ref{eq74}), which are denoted as 
$n_{s0}$, ${\rm {\bf j}}_{ps0}^{(T)} $, $\Delta_{s0} $ and $\Delta_{s0}^{\ast }$, 
correspond to those calculated by means of Eq. (\ref{eq73}). That is,
\begin{equation}
\label{eq76}
J_{s0} =J_{s}^{v_{s} -\mu ,\,{\rm {\bf A}}_{s} ,D_{s} ,D_{s}^{\ast } } 
\left[ {n_{s0} ,{\rm {\bf j}}_{ps0}^{(T)} ,\Delta_{s0} ,\Delta_{s0}^{\ast 
} } \right],
\end{equation}
with
\begin{eqnarray}
\label{eq77}
&&n_{s0} ({\rm {\bf r}})=\mbox{Tr}\left\{ {\hat{{\rho }}_{s0} \hat{{n}}({\rm 
{\bf r}})} \right\}\,, 
\\
\label{eq78}
&&{\rm {\bf j}}_{ps0}^{(T)} ({\rm {\bf r}})=\mbox{Tr}\left\{ {\hat{{\rho 
}}_{s0} {\rm {\bf \hat{{j}}}}_{p}^{(T)} ({\rm {\bf r}})} \right\},
\\
\label{eq79}
&&\Delta_{s0} ({\rm {\bf r}}\zeta ,{\rm {\bf {r}'}}{\zeta 
}')=\mbox{Tr}\left\{ {\hat{{\rho }}_{s0} \hat{{\Delta }}({\rm {\bf r}}\zeta 
,{\rm {\bf {r}'}}{\zeta }')} \right\},
\\
\label{eq80}
&&\Delta_{s0}^{\ast }({\rm {\bf r}}\zeta ,{\rm {\bf {r}'}}{\zeta 
}')=\mbox{Tr}\left\{ {\hat{{\rho }}_{s0} \hat{{\Delta }}^{\dag }({\rm {\bf 
r}}\zeta ,{\rm {\bf {r}'}}{\zeta }')} \right\}.
\end{eqnarray}
This is the variational principle with respect to the densities in the 
reference system. As mentioned later, the effective mean-field potentials 
are determined so that $(n_{s0} ,\,\,{\rm {\bf j}}_{ps0}^{(T)} ,\,\,\Delta 
_{s0} ,\,\,\Delta_{s0}^{\ast } )$ coincide with the equilibrium densities 
of the real system, i.e., ($n_{0} ,\,\,{\rm {\bf j}}_{p0}^{(T)} ,\,\,\Delta 
_{0} ,\,\,\Delta_{0}^{\ast } )$ given by Eqs. (\ref{eq33}) -- (\ref{eq36}).

The one-to-one correspondence between the statistical operator $\hat{{\rho 
}}_{s0} $ and densities $(n_{s0} ,\,\,{\rm {\bf j}}_{ps0}^{(T)} ,\,\,\Delta 
_{s0} ,\,\,\Delta_{s0}^{\ast } )$ can be proven similarly to the case of 
the real system (Sec. III). Namely, if the minimizing statistical operator 
of Eq. (\ref{eq71}) that yields the prescribed densities ($n_{s0} ,{\rm {\bf 
j}}_{ps0}^{(T)} ,\Delta_{s0} ,\Delta_{s0}^{\ast } )$ is denoted as 
$\hat{{\rho }}_{s,\,\min } \left[ {n_{s0} ,{\rm {\bf j}}_{ps0}^{(T)} ,\Delta 
_{s0} ,\Delta_{s0}^{\ast }} \right]$, then we can obtain
\begin{equation}
\label{eq81}
\hat{{\rho }}_{s,\,\min } \left[ {n_{s0} ,{\rm {\bf j}}_{ps0}^{(T)} ,\Delta 
_{s0} ,\Delta_{s0}^{\ast }} \right]=\hat{{\rho }}_{s0} .
\end{equation}
Thus, the one-to-one correspondence between $\hat{{\rho }}_{s0} $ and 
densities ($n_{s0} ,\,\,{\rm {\bf j}}_{ps0}^{(T)} ,\,\,\Delta_{s0} 
,\,\,\Delta_{s0}^{\ast } )$ holds in the reference system.
%
%********************* Sec. V B ******************************************
\subsection{Effective mean-field potentials}
\label{secV-B}
According to the HK theorem for the reference system, the functional 
$J_{s}^{v_{s} -\mu ,\,{\rm {\bf A}}_{s} ,D_{s} ,D_{s}^{\ast } } \left[ 
{n,{\rm {\bf j}}_{p}^{(T)} ,\Delta ,\Delta^{\ast }} \right]$ takes the 
minimum value at densities $(n_{s0} ,{\rm {\bf j}}_{ps0}^{(T)} ,\Delta_{s0} 
,\Delta_{s0}^{\ast } )$. Namely, substituting Eq. (\ref{eq75}) into 
\begin{eqnarray}
\label{eq82}
&&J_{s}^{v_{s} -\mu ,\,{\rm {\bf A}}_{s} ,D_{s} ,D_{s}^{\ast } } \left[ 
{n_{s0} +\delta n,{\rm {\bf j}}_{ps0}^{(T)} +\delta {\rm {\bf j}}_{p}^{(T)} 
,\Delta_{s0} +\delta \Delta ,\Delta_{s0}^{\ast } +\delta \Delta^{\ast }} 
\right] \nonumber \\
&&-J_{s}^{v_{s} -\mu ,\,{\rm {\bf A}}_{s} ,D_{s} ,D_{s}^{\ast } } 
\left[ {n_{s0} ,{\rm {\bf j}}_{ps0}^{(T)} ,\Delta_{s0} ,\Delta_{s0}^{\ast 
} } \right]=0,
\end{eqnarray}
leads to
\begin{eqnarray}
\label{eq83}
&&\left. {{\frac{\delta F_{s} \left[ {n,{\rm {\bf j}}_{p}^{(T)} ,\Delta 
,\Delta^{\ast }} \right]}{\delta n({\rm {\bf r}})}}} 
\right|_{{\begin{array}{l}
 \vspace{-3mm}
 \scriptstyle{n=n_{s0}} \\ \vspace{-3mm}
 \scriptstyle{{\rm {\bf j}}_{p}^{(T)} ={\rm {\bf j}}_{ps0}^{(T)}} \\ \vspace{-3mm}
 \scriptstyle{\Delta =\Delta_{s0}} \\ \vspace{-3mm}
 \scriptstyle{\Delta^{\ast }=\Delta_{s0}^{\ast }} \\ \vspace{-3mm}
 \end{array}}} +\left\{ {v_{s} ({\rm {\bf r}})-\mu } 
\right\}+{\frac{e^{2}}{2m}}{\rm {\bf A}}_{s} ({\rm {\bf r}})^{2}=0,
\\
\label{eq84}
&&\left. {{\frac{\delta F_{s} \left[ {n,{\rm {\bf j}}_{p}^{(T)} ,\Delta 
,\Delta^{\ast }} \right]}{\delta {\rm {\bf j}}_{p}^{(T)} ({\rm {\bf r}})}}} 
\right|_{{\begin{array}{l}
 \vspace{-3mm}
 \scriptstyle{n=n_{s0}} \\  \vspace{-3mm}
 \scriptstyle{{\rm {\bf j}}_{p}^{(T)} ={\rm {\bf j}}_{ps0}^{(T)}} \\  \vspace{-3mm}
 \scriptstyle{ \Delta =\Delta_{s0}} \\  \vspace{-3mm}
 \scriptstyle{\Delta^{\ast }=\Delta_{s0}^{\ast }} \\  \vspace{-3mm}
 \end{array}}} +e{\rm {\bf A}}_{s} ({\rm {\bf r}})=0,
\\
\label{eq85}
&&\left. {{\frac{\delta F_{s} \left[ {n,{\rm {\bf j}}_{p}^{(T)} ,\Delta 
,\Delta^{\ast }} \right]}{\delta \Delta ({\rm {\bf r}}\zeta ,{\rm {\bf 
{r}'}}{\zeta }')}}} \right|_{{\begin{array}{l}
 \vspace{-3mm}
 \scriptstyle{n=n_{s0}} \\ \vspace{-3mm}
 \scriptstyle{{\rm {\bf j}}_{p}^{(T)} ={\rm {\bf j}}_{ps0}^{(T)}} \\ \vspace{-3mm}
 \scriptstyle{\Delta =\Delta_{s0}} \\ \vspace{-3mm}
 \scriptstyle{\Delta^{\ast }=\Delta_{s0}^{\ast }} \\ \vspace{-3mm}
 \end{array}}} +D_{s}^{\ast } ({\rm {\bf r}}\zeta, {\rm {\bf {r}'}}{\zeta 
}')=0,
\\
\label{eq86}
&&\left. {{\frac{\delta F_{s} \left[ {n,{\rm {\bf j}}_{p}^{(T)} ,\Delta 
,\Delta^{\ast }} \right]}{\delta \Delta^{\ast }({\rm {\bf r}}\zeta ,{\rm 
{\bf {r}'}}{\zeta }')}}} \right|_{{\begin{array}{l}
 \vspace{-3mm}
 \scriptstyle{n=n_{s0}} \\ \vspace{-3mm}
 \scriptstyle{{\rm {\bf j}}_{p}^{(T)} ={\rm {\bf j}}_{ps0}^{(T)}} \\ \vspace{-3mm}
 \scriptstyle{\Delta =\Delta_{s0}} \\ \vspace{-3mm}
 \scriptstyle{\Delta^{\ast }=\Delta_{s0}^{\ast }} \\ \vspace{-3mm}
 \end{array}}} +D_{s} ({\rm {\bf r}}\zeta, {\rm {\bf {r}'}}{\zeta }')=0.
\end{eqnarray}

Using $F_{s} \left[ {n,{\rm {\bf j}}_{p}^{(T)} ,\Delta ,\Delta^{\ast }} 
\right]$ and the classical Coulomb interactions between electrons, the 
universal energy functional of the real system, i.e., $F\left[ {n,{\rm {\bf 
j}}_{p}^{(T)} ,\Delta ,\Delta^{\ast }} \right]$, is formally decomposed 
into the following form: 
\begin{equation}
\label{eq87}
F\left[ {n,{\rm {\bf j}}_{p}^{(T)} ,\Delta ,\Delta^{\ast }} \right]=F_{s} 
\left[ {n,{\rm {\bf j}}_{p}^{(T)} ,\Delta ,\Delta^{\ast }} 
\right]+{\frac{e^{2}}{2}}\int\!\!\!\int {{\frac{n({\rm {\bf r}})n({\rm {\bf 
{r}'}})}{\left| {{\rm {\bf r}}-{\rm {\bf {r}'}}} \right|}}d^{3}rd^{3}{r}'} 
+F_{xc} \left[ {n,{\rm {\bf j}}_{p}^{(T)} ,\Delta ,\Delta^{\ast }} 
\right],
\end{equation}
where $F_{xc} \left[ {n,{\rm {\bf j}}_{p}^{(T)} ,\Delta ,\Delta^{\ast }} \right]$ 
is the xc energy functional which contains the quantum effects of 
the electron-electron interaction, the difference of the kinetic energy 
between the real and reference systems, and the difference of the entropy 
between these two systems. Substituting Eq. (\ref{eq87}) into Eq. (\ref{eq29}), and using 
the HK theorem given by Eq. (\ref{eq32}), then we have 
\begin{eqnarray}
\label{eq88}
&&\left. {{\frac{\delta F_{s} \left[ {n,{\rm {\bf j}}_{p}^{(T)} ,\Delta 
,\Delta^{\ast }} \right]}{\delta n({\rm {\bf r}})}}} 
\right|_{{\begin{array}{l}
 \vspace{-3mm}
 \scriptstyle{n=n_{0}} \\ \vspace{-3mm}
 \scriptstyle{{\rm {\bf j}}_{p}^{(T)} ={\rm {\bf j}}_{p0}^{(T)}} \\ \vspace{-3mm}
 \scriptstyle{\Delta =\Delta_{0}} \\ \vspace{-3mm}
 \scriptstyle{\Delta^{\ast }=\Delta_{0}^{\ast }} \\ \vspace{-3mm}
 \end{array}}} +e^{2}\int {{\frac{n_{0} ({\rm {\bf {r}'}})}{\left| {{\rm 
{\bf r}}-{\rm {\bf {r}'}}} \right|}}d^{3}{r}'} 
\nonumber \\
&&+\left. {{\frac{\delta F_{xc} 
\left[ {n,{\rm {\bf j}}_{p}^{(T)} ,\Delta ,\Delta^{\ast }} \right]}{\delta 
n({\rm {\bf r}})}}} \right|_{{\begin{array}{l}
 \vspace{-3mm}
 \scriptstyle{n=n_{0}} \\ \vspace{-3mm}
 \scriptstyle{{\rm {\bf j}}_{p}^{(T)} ={\rm {\bf j}}_{p0}^{(T)}} \\ \vspace{-3mm}
 \scriptstyle{\Delta =\Delta_{0}} \\ \vspace{-3mm}
 \scriptstyle{\Delta^{\ast }=\Delta_{0}^{\ast }} \\ \vspace{-3mm}
 \end{array}}}  
+\left\{ {v_{given} ({\rm {\bf r}})-\mu } \right\}+{\frac{e^{2}}{2m}}{\rm 
{\bf A}}_{given} ({\rm {\bf r}})^{2}=0, 
\\ %******************
\label{eq89}
&&\left. {{\frac{\delta F_{s} \left[ {n,{\rm {\bf j}}_{p}^{(T)} ,\Delta 
,\Delta^{\ast }} \right]}{\delta {\rm {\bf j}}_{p}^{(T)} ({\rm {\bf r}})}}} 
\right|_{{\begin{array}{l}
 \vspace{-3mm}
 \scriptstyle{n=n_{0}} \\ \vspace{-3mm}
 \scriptstyle{{\rm {\bf j}}_{p}^{(T)} ={\rm {\bf j}}_{p0}^{(T)} } \\ \vspace{-3mm}
 \scriptstyle{\Delta =\Delta_{0} } \\ \vspace{-3mm}
 \scriptstyle{\Delta^{\ast }=\Delta_{0}^{\ast } } \\ \vspace{-3mm}
 \end{array}}} +\left. {{\frac{\delta F_{xc} \left[ {n,{\rm {\bf 
j}}_{p}^{(T)} ,\Delta ,\Delta^{\ast }} \right]}{\delta {\rm {\bf 
j}}_{p}^{(T)} ({\rm {\bf r}})}}} \right|_{{\begin{array}{l}
 \vspace{-3mm} 
 \scriptstyle{n=n_{0} } \\ \vspace{-3mm} 
 \scriptstyle{{\rm {\bf j}}_{p}^{(T)} ={\rm {\bf j}}_{p0}^{(T)}} \\ \vspace{-3mm}
 \scriptstyle{\Delta =\Delta_{0}} \\ \vspace{-3mm} 
 \scriptstyle{\Delta^{\ast }=\Delta_{0}^{\ast }} \\ \vspace{-3mm} 
 \end{array}}} +e{\rm {\bf A}}_{given} ({\rm {\bf r}})=0,\vspace{10mm} 
\\ %******************
\label{eq90}
&&\left. {{\frac{\delta F_{s} \left[ {n,{\rm {\bf j}}_{p}^{(T)} ,\Delta 
,\Delta^{\ast }} \right]}{\delta \Delta ({\rm {\bf r}}\zeta ,{\rm {\bf 
{r}'}}{\zeta }')}}} \right|_{{\begin{array}{l}
 \vspace{-3mm}
 \scriptstyle{n=n_{0}} \\ \vspace{-3mm}
 \scriptstyle{{\rm {\bf j}}_{p}^{(T)} ={\rm {\bf j}}_{p0}^{(T)} } \\ \vspace{-3mm}
 \scriptstyle{\Delta =\Delta_{0} } \\ \vspace{-3mm}
 \scriptstyle{\Delta^{\ast }=\Delta_{0}^{\ast } } \\ \vspace{-3mm}
 \end{array}}} +\left. {{\frac{\delta F_{xc} \left[ {n,{\rm {\bf 
j}}_{p}^{(T)} ,\Delta ,\Delta^{\ast }} \right]}{\delta \Delta ({\rm {\bf 
r}}\zeta ,{\rm {\bf {r}'}}{\zeta }')}}} \right|_{{\begin{array}{l}
 \vspace{-3mm}
 \scriptstyle{n=n_{0}} \\ \vspace{-3mm} 
 \scriptstyle{{\rm {\bf j}}_{p}^{(T)} ={\rm {\bf j}}_{p0}^{(T)}} \\ \vspace{-3mm}  
 \scriptstyle{\Delta =\Delta_{0} } \\ \vspace{-3mm} 
 \scriptstyle{\Delta^{\ast }=\Delta_{0}^{\ast }} \\ \vspace{-3mm} 
 \end{array}}} =0,
\\ %******************
\label{eq91}
&&\left. {{\frac{\delta F_{s} \left[ {n,{\rm {\bf j}}_{p}^{(T)} ,\Delta 
,\Delta^{\ast }} \right]}{\delta \Delta^{\ast }({\rm {\bf r}}\zeta ,{\rm 
{\bf {r}'}}{\zeta }')}}} \right|_{{\begin{array}{l}
 \vspace{-3mm}
 \scriptstyle{n=n_{0}} \\ \vspace{-3mm} 
 \scriptstyle{{\rm {\bf j}}_{p}^{(T)} ={\rm {\bf j}}_{p0}^{(T)} } \\ \vspace{-3mm}
 \scriptstyle{\Delta =\Delta_{0}} \\ \vspace{-3mm}
 \scriptstyle{\Delta^{\ast }=\Delta_{0}^{\ast }} \\ \vspace{-3mm}
 \end{array}}} +\left. {{\frac{\delta F_{xc} \left[ {n,{\rm {\bf 
j}}_{p}^{(T)} ,\Delta ,\Delta^{\ast }} \right]}{\delta \Delta^{\ast }({\rm 
{\bf r}}\zeta ,{\rm {\bf {r}'}}{\zeta }')}}} \right|_{{\begin{array}{l}
 \vspace{-3mm}
 \scriptstyle{n=n_{0} } \\ \vspace{-3mm} 
 \scriptstyle{{\rm {\bf j}}_{p}^{(T)} ={\rm {\bf j}}_{p0}^{(T)} } \\ \vspace{-3mm}
 \scriptstyle{\Delta =\Delta_{0} } \\ \vspace{-3mm}
 \scriptstyle{\Delta^{\ast }=\Delta_{0}^{\ast }} \\ \vspace{-3mm}
 \end{array}}} =0.
\end{eqnarray}
Equations (\ref{eq83}) -- (\ref{eq86}) are satisfied by the densities of the 
variationally-minimum point of the reference system, i.e., ($n_{s0} ,{\rm 
{\bf j}}_{ps0}^{(T)} ,\Delta_{s0} ,\Delta_{s0}^{\ast } )$, while Eqs. (\ref{eq88}) 
-- (\ref{eq91}) are satisfied by the correct densities of the real system, i.e., 
($n_{0} ,{\rm {\bf j}}_{p0}^{(T)} ,\Delta_{0} ,\Delta_{0}^{\ast } )$. If 
Eqs. (\ref{eq83}) -- (\ref{eq86}) coincide with Eqs. (\ref{eq88}) -- (\ref{eq91}), respectively, then Eqs. 
(\ref{eq83}) -- (\ref{eq86}) can be regarded as the equations that are satisfied by the 
correct densities ($n_{0} ,{\rm {\bf j}}_{p0}^{(T)} ,\Delta_{0} ,\Delta 
_{0}^{\ast } )$. Specifically the following effective mean-filed potentials 
can reproduce the correct densities as the densities of the 
variationally-minimum point of the reference system:
\begin{eqnarray}
\label{eq92}
&&v_{s} ({\rm {\bf r}})\!=\!v_{given} ({\rm {\bf r}})\!+\!e^{2}\!\!\int \!\!{{\frac{n_{0} 
({\rm {\bf {r}'}})}{\left| {{\rm {\bf r}}-{\rm {\bf {r}'}}} 
\right|}}d^{3}{r}'} \!+\!\left. {{\frac{\delta F_{xc} [n,{\rm {\bf j}}_{p}^{(T)} 
,\Delta ,\Delta^{\ast }]}{\delta n({\rm {\bf r}})}}} 
\right|_{{\begin{array}{l}
 \vspace{-3mm}
 \scriptstyle{n=n_{0}} \\ \vspace{-3mm} 
 \scriptstyle{j_{p}^{(T)} =j_{p0}^{(T)}} \\ \vspace{-3mm} 
 \scriptstyle{\Delta =\Delta_{0}} \\ \vspace{-3mm} 
 \scriptstyle{\Delta^{\ast }=\Delta_{0}^{\ast } } \\ \vspace{-3mm}
 \end{array}}} 
\!\!\!\!\!\!\!\!\!+\!{\frac{e^{2}}{2m}}\left\{ \!{{\rm {\bf A}}_{given}^{2} ({\rm 
{\bf r}})\!-\!{\rm {\bf A}}_{s}^{2} \!({\rm {\bf r}})} \right\}\!\!,\ \ \ \ \ \ \ 
\\
\label{eq93}
&&{\rm {\bf A}}_{s} ({\rm {\bf r}})={\rm {\bf A}}_{given} ({\rm {\bf 
r}})+\left. {{\frac{1}{e}}{\frac{\delta F_{xc} [n,{\rm {\bf j}}_{p}^{(T)} 
,\Delta ,\Delta^{\ast }]}{\delta {\rm {\bf j}}_{p}^{(T)} ({\rm {\bf r}})}}} 
\right|_{{\begin{array}{l}
 \vspace{-3mm}
 \scriptstyle{n=n_{0} } \\ \vspace{-3mm}
 \scriptstyle{j_{p}^{(T)} =j_{p0}^{(T)} } \\ \vspace{-3mm}
 \scriptstyle{\Delta =\Delta_{0}} \\ \vspace{-3mm}
 \scriptstyle{\Delta^{\ast }=\Delta_{0}^{\ast }} \\ \vspace{-3mm} 
 \end{array}}} ,
\\
\label{eq94}
&&D_{s}^{\ast } ({\rm {\bf r}}\zeta ,{\rm {\bf {r}'}}{\zeta }')=\left. 
{{\frac{\delta F_{xc} [n,{\rm {\bf j}}_{p}^{(T)} ,\Delta ,\Delta^{\ast 
}]}{\delta \Delta ({\rm {\bf r}}\zeta ,{\rm {\bf {r}'}}{\zeta }')}}} 
\right|_{{\begin{array}{l}
 \vspace{-3mm}
 \scriptstyle{n=n_{0}} \\ \vspace{-3mm} 
 \scriptstyle{j_{p}^{(T)} =j_{p0}^{(T)} } \\ \vspace{-3mm} 
 \scriptstyle{\Delta =\Delta_{0} } \\ \vspace{-3mm} 
 \scriptstyle{\Delta^{\ast }=\Delta_{0}^{\ast } } \\ \vspace{-3mm}
 \end{array}}} ,
\\
\label{eq95}
&&D_{s} ({\rm {\bf r}}\zeta ,{\rm {\bf {r}'}}{\zeta }')=\left. {{\frac{\delta 
F_{xc} [n,{\rm {\bf j}}_{p}^{(T)} ,\Delta ,\Delta^{\ast }]}{\delta \Delta 
^{\ast }({\rm {\bf r}}\zeta ,{\rm {\bf {r}'}}{\zeta }')}}} 
\right|_{{\begin{array}{l}
 \vspace{-3mm}
 \scriptstyle{n=n_{0}} \\ \vspace{-3mm} 
 \scriptstyle{j_{p}^{(T)} =j_{p0}^{(T)} } \\ \vspace{-3mm}
 \scriptstyle{\Delta =\Delta_{0} } \\ \vspace{-3mm} 
 \scriptstyle{\Delta^{\ast }=\Delta_{0}^{\ast }  } \\ \vspace{-3mm}  
 \end{array}}} .
\end{eqnarray}
The key point to determine the effective mean-field potentials is that the 
reference system is prepared so that the densities ($n_{s0} ,{\rm {\bf 
j}}_{ps0}^{(T)} ,\Delta_{s0} ,\Delta_{s0}^{\ast } )$ coincide with the 
correct densities ($n_{0} ,{\rm {\bf j}}_{p0}^{(T)} ,\Delta_{0} ,\Delta 
_{0}^{\ast } )$ of the real system.
%
%********************* Sec. V C ******************************************
\subsection{Basic variables in the reference system}
\label{secV-C}
The basic variables are expressed by using the eigenvalues and 
eigenfunctions of the BdG-KS equation. By solving the BdG-KS equation, and 
using the resultant solutions, the Hamiltonian of the reference system is 
written in the diagonalized form given by Eq. (\ref{eq47}). Using Eq. (\ref{eq47}), the 
basic variables of the reference system can be calculated via Eqs. (\ref{eq73}) and 
(\ref{eq77}) -- (\ref{eq80}). We have
\begin{eqnarray}
\label{eq96}
&&n_{0} ({\rm {\bf r}})=\sum\limits_i {f(E_{i} )\int {\left| {u_{i} ({\rm {\bf 
r}}\zeta )} \right|} }^{2}d\zeta +\sum\limits_i {\left\{ {1-f(E_{i} )} 
\right\}\int {\left| {v_{i} ({\rm {\bf r}}\zeta )} \right|} }^{2}d\zeta ,
\\
\label{eq97}
&&{\rm {\bf j}}_{p0}^{(T)} ({\rm {\bf r}})={\frac{\hbar }{i4\pi 
m}}\sum\limits_i {\left[ {f(E_{i} )\int {\left\{ {\nabla_{{\rm {\bf {r}'}}} 
u_{i}^{\ast } ({\rm {\bf {r}'}}{\zeta }')\times \nabla_{{\rm {\bf {r}'}}} 
u_{i} ({\rm {\bf {r}'}}{\zeta }')} \right\}\times {\frac{{\rm {\bf r}}-{\rm 
{\bf {r}'}}}{\left| {{\rm {\bf r}}-{\rm {\bf {r}'}}} 
\right|^{3}}}d^{3}{r}'d\zeta } } \right.} \nonumber \\ 
&&\hspace{20mm}
\left. {+\left\{ {1-f(E_{i} )} \right\}\int {\left\{ {\nabla_{{\rm {\bf 
{r}'}}} v_{i}^{\ast } ({\rm {\bf {r}'}}{\zeta }')\times \nabla_{{\rm {\bf 
{r}'}}} v_{i} ({\rm {\bf {r}'}}{\zeta }')} \right\}\times {\frac{{\rm {\bf 
r}}-{\rm {\bf {r}'}}}{\left| {{\rm {\bf r}}-{\rm {\bf {r}'}}} 
\right|^{3}}}d^{3}{r}'d\zeta } } \right], 
\\
\label{eq98}
&&\Delta_{0} ({\rm {\bf r}}\zeta ,{\rm {\bf {r}'}}{\zeta }')=\sum\limits_i 
{\left[ {u_{i} ({\rm {\bf {r}'}}{\zeta }')v_{i} ({\rm {\bf r}}\zeta )\left\{ 
{1-f(E_{i} )} \right\}+v_{i} ({\rm {\bf {r}'}}{\zeta }')u_{i} ({\rm {\bf 
r}}\zeta )f(E_{i} )} \right]} ,
\\
\label{eq99}
&&\Delta_{0}^{\ast } ({\rm {\bf r}}\zeta ,{\rm {\bf {r}'}}{\zeta 
}')=\sum\limits_i {\left[ {u_{i}^{\ast } ({\rm {\bf {r}'}}{\zeta 
}')v_{i}^{\ast } ({\rm {\bf r}}\zeta )\left\{ {1-f(E_{i} )} 
\right\}+v_{i}^{\ast } ({\rm {\bf {r}'}}{\zeta }')u_{i}^{\ast } ({\rm {\bf 
r}}\zeta )f(E_{i} )} \right]} ,
\end{eqnarray}
where $f(E_{i} )$ is the fermi distribution function given by
\begin{equation}
\label{eq100}
f(E_{i} )={\frac{1}{e^{\beta E_{i} }+1}}.
\end{equation}
%

%********************* Sec. V D ******************************************
\subsection{Calculation procedure}
\label{secV-D}
The effective mean-field potentials given by Eqs. (\ref{eq92}) -- (\ref{eq95}) are dependent 
on the correct densities ($n_{0} ,{\rm {\bf j}}_{p0}^{(T)} ,\Delta_{0} 
,\Delta_{0}^{\ast } )$, and should be determined in a self-consistent way. 
The concrete steps of calculating them are as follows: (i) the input set of 
densities ($n_{0} ,{\rm {\bf j}}_{p0}^{(T)} ,\Delta_{0} ,\Delta_{0}^{\ast 
} )$ is tentatively given, (ii) using Eqs. (\ref{eq92}) -- (\ref{eq95}), the effective 
mean-field potentials are calculated, (iii) solving the BdG-KS equation, and 
substituting these solutions into Eqs. (\ref{eq96}) -- (\ref{eq99}), the new set of 
densities ($n_{0} ,{\rm {\bf j}}_{p0}^{(T)} ,\Delta_{0} ,\Delta_{0}^{\ast 
} )$ is obtained, (iv) comparing this set with the input set, and if these 
are not consistent with each other within some accuracy then the 
calculations are restarted from (i) with changing the input set. The 
calculations are repeated until the self-consistency mentioned above is 
attained. 

From the HK theorem, the densities thus obtained correspond to the 
equilibrium densities for the beforehand-given electromagnetic fields 
$v_{given} ({\rm {\bf r}})$ and ${\rm {\bf A}}_{given} ({\rm {\bf r}})$. The 
field $v_{given} ({\rm {\bf r}})$ originates from the charged particles 
positioned outside the system. On the other hand, ${\rm {\bf A}}_{given} 
({\rm {\bf r}})$ is the potential that electrons of the system feel at the 
position ${\rm {\bf r}}$, and should be consistent with the densities $n_{0} 
({\rm {\bf r}})$ and ${\rm {\bf j}}_{p0}^{(T)} ({\rm {\bf r}})$. Namely, the 
beforehand-given field ${\rm {\bf A}}_{given} ({\rm {\bf r}})$ consists of 
the field caused by the charged particles positioned outside the system, and 
the field caused by the charged particles inside the system. The former 
field is the completely fixed one, while the later field should be 
determined by using the densities $n_{0} ({\rm {\bf r}})$ and ${\rm {\bf 
j}}_{p0}^{(T)} ({\rm {\bf r}})$ via the microscopic Maxwell equation 
\cite{59,60}. Therefore, in addition to the self-consistent calculation loop for 
the densities $(n_{0} ,{\rm {\bf j}}_{p0}^{(T)} ,\Delta_{0} ,\Delta 
_{0}^{\ast } )$, which is explained above, we need further calculation loop 
to keep the consistency between the beforehand-given electromagnetic field 
${\rm {\bf A}}_{given} ({\rm {\bf r}})$ and the densities $\left( {n_{0} 
({\rm {\bf r}}),\,\,{\rm {\bf j}}_{p0}^{(T)} ({\rm {\bf r}})} \right)$. The 
concrete steps of the calculations are as follows: (I) $v_{given} ({\rm {\bf 
r}})$ is determined in accordance with the charged particles outside the 
system, while ${\rm {\bf A}}_{given} ({\rm {\bf r}})$ is tentatively given, 
(II) using these fields, the corresponding densities $(n_{0} ,{\rm {\bf 
j}}_{p0}^{(T)} ,\Delta_{0} ,\Delta_{0}^{\ast } )$ are obtained via the 
calculation loop denoted as (i) -- (iv), (III) substituting these densities 
into the microscopic Maxwell equation \cite{61}, the electromagnetic fields 
caused by the charged particles inside the system are obtained, (IV) adding 
these fields to the fixed external fields, the new set of the 
electromagnetic fields are obtained, (V) taking ${\rm {\bf A}}_{given} ({\rm 
{\bf r}})$ among the thus-obtained electromagnetic fields, and comparing it 
with the input one, and if these are not consistent with each other within 
some accuracy then the calculations are restarted from (I) with changing the 
input field. The calculations are repeated until the self-consistency is 
attained.

%
%********************* Sec. VI ******************************************
\section{Exchange-Correlation energy functional}
\label{secVI}
In the development of the DFT-based or ECS-based theory, the following 
issues have to be performed. Both of them are indispensable so that the 
theory works well:

\begin{enumerate}
\renewcommand{\labelenumi}{(\alph{enumi})}
\item The proof of the HK theorem and the derivation of the KS equation,
\item Development of the approximate form of the xc energy functional.
\end{enumerate}
We have performed (a) in the preceding sections. In this section, we shall 
tackle with (b) in two ways. One is the derivation of the rigorous 
expression for the xc energy functional with the use of the 
coupling-constant integration. This kind of expression seems to be useful 
for the development of the approximate forms in the future. Actually, in the 
conventional DFT \cite{23,24}, the local density approximation (LDA) for the xc 
energy functional was developed on the basis of the rigorous expression with 
the coupling-constant integration \cite{24,62}. The other is a proposal of the 
practically useful form of the xc energy functional with the use of the 
approximate solutions of the BdG-KS equation. (Eq. (\ref{eq53})). Specifically, it 
is devised so that the solution of the BdG-KS equation has the 
superconducting energy gap which is consistent with the attractive 
interaction appeared in the Hamiltonian of the real system.
%
%********************* Sec. VI A ******************************************
\subsection{Rigorous expression with the use of the coupling-constant integration}
\label{secVI-A}
The Hamiltonian of the real system is given by Eq. (\ref{eq5}), while that of the 
reference system is given by Eq. (\ref{eq66}). Here we shall introduce the system 
which is expressed by the Hamiltonian scaled with the coupling constant $\xi$:
\begin{equation}
\label{eq101}
\hat{{H}}_{\xi } =\hat{{T}}+\xi \left( {\hat{{W}}_{1} +\hat{{W}}_{2} } 
\right)+\hat{{U}}_{\xi } ,
\end{equation}
where $\hat{{U}}_{\xi } $ is chosen to be the external potential which makes 
the equilibrium densities of the system $\hat{{H}}_{\xi } $ be coincident 
with the correct densities ($n_{0} ,{\rm {\bf j}}_{p0}^{(T)} ,\Delta_{0} 
,\Delta_{0}^{\ast } )$. Let $\hat{{U}}_{\xi } $ be expressed as 
\begin{eqnarray}
\label{eq102}
\hat{{U}}_{\xi } 
&=&
\int {v_{\xi } ({\rm {\bf r}})\hat{{n}}({\rm {\bf r}})} 
d^{3}r+e\int {{\rm {\bf A}}_{\xi } ({\rm {\bf r}})\cdot {\rm {\bf 
\hat{{j}}}}_{p}^{(T)} ({\rm {\bf r}})d^{3}r+} {\frac{e^{2}}{2m}}\int {{\rm 
{\bf A}}_{\xi } ({\rm {\bf r}})^{2}\hat{{n}}({\rm {\bf r}})d^{3}r} \nonumber \\ 
&&+
\int\!\!\!\int {D_{\xi }^{\ast } ({\rm {\bf r}}\zeta, {\rm {\bf 
{r}'}}{\zeta }')\hat{{\Delta }}({\rm {\bf r}}\zeta, {\rm {\bf {r}'}}{\zeta 
}')d^{3}rd\zeta d^{3}{r}'d{\zeta }'} \nonumber \\ 
&&+
\int\!\!\!\int {D_{\xi } ({\rm {\bf r}}\zeta, {\rm {\bf {r}'}}{\zeta 
}')\hat{{\Delta }}^{\dag }({\rm {\bf r}}\zeta, {\rm {\bf {r}'}}{\zeta 
}')d^{3}rd\zeta d^{3}{r}'d{\zeta }'} . 
\end{eqnarray}
In the special cases of $\xi =1$ and $\xi =0$, $\hat{{U}}_{\xi } $ is given by
\begin{equation}
\label{eq103}
\hat{{U}}_{\xi } =\left\{ {{\begin{array}{*{20}c}
 {\hat{{V}}_{1} +\hat{{V}}_{2} +\hat{{V}}_{3} 
\hspace{20mm} \mbox{for}\,\,\,\xi =1,} \hfill \\
 {\hat{{V}}_{1s} +\hat{{V}}_{2s} +\hat{{V}}_{3s} +\,\hat{{V}}_{Ds} 
\,\,\,\,\,\,\mbox{for}\,\,\,\xi =0,} \hfill \\
\end{array} }} \right.
\end{equation}
respectively. We shall define the following universal energy functional
\begin{eqnarray}
\label{eq104}
F_{\xi } \left[ {n,\,{\rm {\bf j}}_{p}^{(T)} ,\,\Delta ,\Delta^{\ast }} 
\right]
&=&
\mathop{\mbox{Min}}\limits_{\hat{{\rho }}\to n,\,{\rm {\bf j}}_{p}^{(T)} 
,\,\Delta ,\Delta^{\ast }} \mbox{Tr}\left[ {\hat{{\rho }}\left\{ 
{\hat{{T}}+\xi \left( {\hat{{W}}_{1} +\hat{{W}}_{2} } \right)} 
\right\}+{\frac{1}{\beta }}\hat{{\rho }}\ln \hat{{\rho }}} \right] \nonumber \\ 
&=&\mbox{Tr}\left[ {\hat{{\rho }}_{\xi } \left[ {n,\,{\rm {\bf j}}_{p}^{(T)} 
,\,\Delta ,\Delta^{\ast }} \right]\left\{ {\hat{{T}}+\xi \left( 
{\hat{{W}}_{1} +\hat{{W}}_{2} } \right)} \right\}} \right.  \nonumber \\ 
&& \left. {+{\frac{1}{\beta }}\hat{{\rho }}_{\xi } \left[ {n,\,{\rm {\bf 
j}}_{p}^{(T)} ,\,\Delta ,\Delta^{\ast }} \right]\ln \hat{{\rho }}_{\xi } 
\left[ {n,\,{\rm {\bf j}}_{p}^{(T)} ,\,\Delta ,\Delta^{\ast }} \right]} 
\right], 
\end{eqnarray}
where the minimizing $\hat{{\rho }}$ is denoted as $\hat{{\rho }}_{\xi } 
\left[ {n,\,{\rm {\bf j}}_{p}^{(T)} ,\,\Delta ,\Delta^{\ast }} \right]$. 
Using the fact that $F_{1} \left[ {n_{0} ,\,{\rm {\bf j}}_{p0}^{(T)} 
,\,\Delta_{0} ,\Delta_{0}^{\ast }} \right]$ and $F_{0} \left[ {n_{0} 
,\,{\rm {\bf j}}_{p0}^{(T)} ,\,\Delta_{0} ,\Delta_{0}^{\ast }} \right]$ 
are coincident with $F\left[ {n_{0} ,\,{\rm {\bf j}}_{p0}^{(T)} ,\,\Delta 
_{0} ,\Delta_{0}^{\ast }} \right]$ and $F_{s} \left[ {n_{0} ,\,{\rm {\bf 
j}}_{p0}^{(T)} ,\,\Delta_{0} ,\Delta_{0}^{\ast }} \right]$, respectively, 
the xc energy functional defined by Eq. (\ref{eq87}) is rewritten as
\begin{equation}
\label{eq105}
F_{xc} \left[ {n_{0} ,\,{\rm {\bf j}}_{p0}^{(T)} ,\,\Delta_{0} ,\Delta_{0} 
^{\ast }} \right]=\int_0^1 {{\frac{dF_{\xi } \left[ {n_{0} ,\,{\rm {\bf 
j}}_{p0}^{(T)} ,\,\Delta_{0} ,\Delta_{0}^{\ast }} \right]}{d\xi }}d\xi } 
-{\frac{e^{2}}{2}}\int\!\!\!\int {{\frac{n_{0} ({\rm {\bf r}})n_{0} ({\rm 
{\bf {r}'}})}{\left| {{\rm {\bf r}}-{\rm {\bf {r}'}}} 
\right|}}d^{3}rd^{3}{r}'} .
\end{equation}

Next we shall consider the integrand ${dF_{\xi } \left[ {n_{0} ,\,{\rm {\bf 
j}}_{p0}^{(T)} ,\,\Delta_{0} ,\Delta_{0}^{\ast }} \right]} 
\mathord{\left/ {\vphantom {{dF_{\xi } \left[ {n_{0} ,\,{\rm {\bf 
j}}_{p0}^{(T)} ,\,\Delta_{0} ,\Delta_{0}^{\ast }} \right]} {d\xi }}} 
\right. \kern-\nulldelimiterspace} {d\xi }$ in Eq. (\ref{eq105}). 
For the 
convenience of discussions, the energy functional of the scaled system, 
which corresponds to Eq. (\ref{eq75}) of the reference system, 
is defined as
\begin{eqnarray}
\label{eq106}
J_{\xi }^{v_{\xi } -\mu ,\,{\rm {\bf A}}_{\xi } ,D_{\xi } ,D_{\xi }^{\ast } 
} \left[ {n,{\rm {\bf j}}_{p}^{(T)} ,\Delta ,\Delta^{\ast }} \right]
&=&
F_{\xi } \left[ {n,{\rm {\bf j}}_{p}^{(T)} ,\Delta ,\Delta^{\ast }} \right]
+\int {\left\{ {v_{\xi } ({\rm {\bf r}})-\mu } \right\}
n({\rm {\bf r}})} d^{3}r \nonumber \\ 
&+&
e\int {{\rm {\bf A}}_{\xi } ({\rm {\bf r}})\cdot {\rm {\bf j}}_{p}^{(T)} 
({\rm {\bf r}})d^{3}r+} {\frac{e^{2}}{2m}}\int {{\rm {\bf A}}_{\xi } ({\rm 
{\bf r}})^{2}n({\rm {\bf r}})d^{3}r} \nonumber \\ 
&+&
\int\!\!\!\int {D_{\xi }^{\ast } ({\rm {\bf r}}\zeta, {\rm {\bf 
{r}'}}{\zeta }')\Delta ({\rm {\bf r}}\zeta, {\rm {\bf {r}'}}{\zeta 
}')d^{3}rd\zeta d^{3}{r}'d{\zeta }'} \nonumber \\ 
&+&
\int\!\!\!\int {D_{\xi } ({\rm {\bf r}}\zeta, {\rm {\bf {r}'}}{\zeta 
}')\Delta^{\ast }({\rm {\bf r}}\zeta, {\rm {\bf {r}'}}{\zeta }')d^{3}rd\zeta 
d^{3}{r}'d{\zeta }'} . 
\end{eqnarray}
Using Eq. (\ref{eq104}), Eq. (\ref{eq106}) is rewritten as
\begin{eqnarray}
\label{eq107}
J_{\xi }^{v_{\xi } -\mu ,\,{\rm {\bf A}}_{\xi } ,D_{\xi } ,D_{\xi }^{\ast } 
} \left[ {n,{\rm {\bf j}}_{p}^{(T)} ,\Delta ,\Delta^{\ast }} \right]
&=&
\mbox{Tr}\left[ {\hat{{\rho }}_{\xi } \left[ {n,\,{\rm {\bf 
j}}_{p}^{(T)} ,\,\Delta ,\Delta^{\ast }} \right]\left\{ {\hat{{H}}_{\xi } 
-\mu \hat{{N}}} \right\}} \right. \nonumber \\ 
&&\left. {+{\frac{1}{\beta }}\hat{{\rho }}_{\xi } \left[ {n,\,{\rm {\bf 
j}}_{p}^{(T)} ,\,\Delta ,\Delta^{\ast }} \right]\ln \hat{{\rho }}_{\xi } 
\left[ {n,\,{\rm {\bf j}}_{p}^{(T)} ,\,\Delta ,\Delta^{\ast }} \right]} 
\right]. 
\end{eqnarray}
Differentiating both sides of Eq. (\ref{eq107}) with respect to $\xi $, 
we have
\begin{eqnarray}
\label{eq108}
{\frac{d}{d\xi }}J_{\xi }^{v_{\xi } -\mu ,\,{\rm {\bf A}}_{\xi } ,D_{\xi } 
D_{\xi }^{\ast } } \left[ {n,{\rm {\bf j}}_{p}^{(T)} ,\Delta ,\Delta^{\ast 
}} \right]
&=&
\mbox{Tr}\left[ {\frac{d\hat{{\rho }}_{\xi } \left[ {n,\,{\rm 
{\bf j}}_{p}^{(T)} ,\,\Delta ,\Delta^{\ast }} \right]}{d\xi }}\left\{ 
{\hat{{H}}_{\xi } -\mu \hat{{N}}} \right\} \right. \nonumber \\
&&+\hat{{\rho }}_{\xi } \left[ 
{n,\,{\rm {\bf j}}_{p}^{(T)} ,\,\Delta ,\Delta^{\ast }} 
\right]{\frac{d\hat{{H}}_{\xi } }{d\xi }} \nonumber \\ 
&&+{\frac{1}{\beta }}{\frac{d\hat{{\rho }}_{\xi } \left[ {n,\,{\rm {\bf 
j}}_{p}^{(T)} ,\,\Delta ,\Delta^{\ast }} \right]}{d\xi }}\ln \hat{{\rho 
}}_{\xi } \left[ {n,\,{\rm {\bf j}}_{p}^{(T)} ,\,\Delta ,\Delta^{\ast }} 
\right] \nonumber \\ 
&& \left. {+{\frac{1}{\beta }}\hat{{\rho }}_{\xi } \left[ {n,\,{\rm {\bf 
j}}_{p}^{(T)} ,\,\Delta ,\Delta^{\ast }} \right]{\frac{d}{d\xi }}\ln 
\hat{{\rho }}_{\xi } \left[ {n,\,{\rm {\bf j}}_{p}^{(T)} ,\,\Delta ,\Delta 
^{\ast }} \right]} \right]. \ \ \ \ \ \ \ 
\end{eqnarray}
Of course, Eq. (\ref{eq108}) holds also for the correct densities ($n_{0} ,\,{\rm 
{\bf j}}_{p0}^{(T)} ,\,\Delta_{0} ,\Delta_{0}^{\ast })$. In that case, we 
can use the HK theorem for the scaled system, the result of which gives the 
relation such that $\hat{{\rho }}_{\xi } \left[ {n_{0} ,\,{\rm {\bf 
j}}_{p0}^{(T)} ,\,\Delta_{0} ,\Delta_{0}^{\ast }} \right]={e^{-\beta 
\left( {\hat{{H}}_{\xi } -\mu \hat{{N}}} \right)}} \mathord{\left/ 
{\vphantom {{e^{-\beta \left( {\hat{{H}}_{\xi } -\mu \hat{{N}}} \right)}} 
{\Xi_{\xi } }}} \right. \kern-\nulldelimiterspace} {\Xi_{\xi } }$ with 
$\Xi_{\xi } =\mbox{Tr}\left\{ {e^{-\beta \left( {\hat{{H}}_{\xi } -\mu 
\hat{{N}}} \right)}} \right\}$. Substituting this relation into Eq. (\ref{eq108}), 
we have
\begin{equation}
\label{eq109}
{\frac{d}{d\xi }}J_{\xi }^{v_{\xi } -\mu ,\,{\rm {\bf A}}_{\xi } ,D_{\xi } 
,D_{\xi }^{\ast } } \left[ {n_{0} ,\,{\rm {\bf j}}_{p0}^{(T)} ,\,\Delta_{0} 
,\Delta_{0}^{\ast }} \right]=\mbox{Tr}\left\{ {\hat{{\rho }}_{\xi } \left[ 
{n_{0} ,\,{\rm {\bf j}}_{p0}^{(T)} ,\,\Delta_{0} ,\Delta_{0}^{\ast }} 
\right]\left( {\hat{{W}}_{1} +\hat{{W}}_{2} +{\frac{d\hat{{U}}_{\xi } }{d\xi 
}}} \right)} \right\}.
\end{equation}
Considering the explicit form of 
$J_{\xi }^{v_{\xi } -\mu ,\,{\rm {\bf A}}_{\xi } ,D_{\xi } ,D_{\xi }^{\ast } } \left[ {n,{\rm 
{\bf j}}_{p}^{(T)} ,\Delta ,\Delta^{\ast }} \right]$ given by Eq. (\ref{eq107}), 
Eq. (\ref{eq109}) can be regarded as the Hellman-Feynman theorem at the finite 
temperature. Namely, the derivative of the statistical operator with respect 
to $\xi $ does not appear but the derivative of the energy operators with 
respect to $\xi $ appears in the RHS of Eq. (\ref{eq109}).

On the other hand, using Eq. (\ref{eq106}), we have
\begin{eqnarray}
\label{eq110}
{\frac{d}{d\xi }}J_{\xi }^{v_{\xi } -\mu ,\,{\rm {\bf A}}_{\xi } ,D_{\xi } 
,D_{\xi }^{\ast } } \left[ {n_{0} ,\,{\rm {\bf j}}_{p0}^{(T)} ,\,\Delta_{0} 
,\Delta_{0}^{\ast }} \right]
&=&
{\frac{dF_{\xi } \left[ {n_{0} ,\,{\rm {\bf 
j}}_{p0}^{(T)} ,\,\Delta_{0} ,\Delta_{0}^{\ast }} \right]_{\xi } }{d\xi }}
\nonumber \\
&&+\mbox{Tr}\left\{ {\hat{{\rho }}_{\xi } \left[ {n_{0} ,\,{\rm {\bf 
j}}_{p0}^{(T)} ,\,\Delta_{0} ,\Delta_{0}^{\ast }} 
\right]{\frac{d\hat{{U}}_{\xi } }{d\xi }}} \right\}
\end{eqnarray}
Comparing Eq. (\ref{eq109}) with Eq. (\ref{eq110}), the expression for ${dF_{\xi } \left[ 
{n_{0} ,\,{\rm {\bf j}}_{p0}^{(T)} ,\,\Delta_{0} ,\Delta_{0}^{\ast }} 
\right]} \mathord{\left/ {\vphantom {{dF_{\xi } \left[ {n_{0} ,\,{\rm {\bf 
j}}_{p0}^{(T)} ,\,\Delta_{0} ,\Delta_{0}^{\ast }} \right]} {d\xi }}} 
\right. \kern-\nulldelimiterspace} {d\xi }$ is obtained. Substituting it 
into Eq. (\ref{eq105}) yields 
\begin{eqnarray}
\label{eq111}
F_{xc} \left[ {n_{0} ,\,{\rm {\bf j}}_{p0}^{(T)} ,\,\Delta_{0} ,\Delta_{0} 
^{\ast }} \right]
&=&
\int_0^1 {\mbox{Tr}\left\{ {\hat{{\rho }}_{\xi } \left[ 
{n_{0} ,\,{\rm {\bf j}}_{p0}^{(T)} ,\,\Delta_{0} ,\Delta_{0}^{\ast }} 
\right]\left( {\hat{{W}}_{1} +\hat{{W}}_{2} } \right)} \right\}d\xi } \nonumber \\
&&-
{\frac{e^{2}}{2}}\int\!\!\!\int {{\frac{n_{0} ({\rm {\bf r}})n_{0} ({\rm 
{\bf {r}'}})}{\left| {{\rm {\bf r}}-{\rm {\bf {r}'}}} 
\right|}}d^{3}rd^{3}{r}'} .
\end{eqnarray}
Equation (\ref{eq111}) is exactly the rigorous expression for the xc energy 
functional of the present theory. 

Let us consider the meaning of the coupling-constant integration in Eq. 
(\ref{eq111}). Rewriting the xc energy functional defined by Eq. (\ref{eq87}) by means of 
the minimizing statistical operators, we have
\begin{eqnarray}
\label{eq112}
F_{xc} \left[ {n_{0} ,\,{\rm {\bf j}}_{p0}^{(T)} ,\,\Delta_{0} ,\Delta 
_{0}^{\ast }} \right]
\!\!&=&\!\!
\mbox{Tr}\left\{ {\hat{{\rho }}_{\min } \left[ {n_{0} 
,\,{\rm {\bf j}}_{p0}^{(T)} ,\,\Delta_{0} ,\Delta_{0}^{\ast }} 
\right] \! \left( \!{\hat{{W}}_{1} \!+\! \hat{{W}}_{2} } \! \right) \!} 
\right\}-{\frac{e^{2}}{2}}\!\int\!\!\!\!\int \!{{\frac{n_{0} ({\rm {\bf r}})n_{0} 
({\rm {\bf {r}'}})}{\left| {{\rm {\bf r}}-{\rm {\bf {r}'}}} 
\right|}}d^{3}rd^{3}{r}'} \nonumber \\ 
&+&
\mbox{Tr}\left\{ {\hat{{\rho }}_{\min } \left[ {n_{0} ,\,{\rm {\bf 
j}}_{p0}^{(T)} ,\,\Delta_{0} ,\Delta_{0}^{\ast }} \right]\hat{{T}}} 
\right\}-\mbox{Tr}\left\{ {\hat{{\rho }}_{s,\,\,\min } \left[ {n_{0} ,\,{\rm 
{\bf j}}_{p0}^{(T)} ,\,\Delta_{0} ,\Delta_{0}^{\ast }} \right]\hat{{T}}} 
\right\}  \nonumber \\ 
&+&{\frac{1}{\beta }}\mbox{Tr}\left\{ {\hat{{\rho }}_{\min } \left[ {n_{0} 
,\,{\rm {\bf j}}_{p0}^{(T)} ,\,\Delta_{0} ,\Delta_{0}^{\ast }} \right]\ln 
\hat{{\rho }}_{\min } \left[ {n_{0} ,\,{\rm {\bf j}}_{p0}^{(T)} ,\,\Delta 
_{0} ,\Delta_{0}^{\ast }} \right]} \right\} \nonumber \\ 
&-&
{\frac{1}{\beta }}\mbox{Tr}\left\{ {\hat{{\rho }}_{s,\,\,\min } \left[ 
{n_{0} ,\,{\rm {\bf j}}_{p0}^{(T)} ,\,\Delta_{0} ,\Delta_{0}^{\ast }} 
\right]\ln \hat{{\rho }}_{s,\,\,\min } \left[ {n_{0} ,\,{\rm {\bf 
j}}_{p0}^{(T)} ,\,\Delta_{0} ,\Delta_{0}^{\ast }} \right]} \right\}, 
\end{eqnarray}
where the first and second terms of the RHS correspond to the xc energy, and 
the third and fourth terms mean the difference of the kinetic energy between 
the real and reference systems, and the fifth and sixth terms mean the 
difference of the entropy between these systems. Comparing this result with 
Eq. (\ref{eq111}), the xc energy functional of the CDFT for the superconductor 
contains not only the xc energy but also the difference of the kinetic 
energy and that of the entropy via the coupling-constant integration \cite{63}. 
%
%********************* Sec. VI B ******************************************
\subsection{An approximate form of the xc energy functional}
\label{secVI-B}
In this subsection, we present the approximate forms of $F_{xc} \left[ 
{n,\,{\rm {\bf j}}_{p}^{(T)} ,\,\Delta ,\Delta^{\ast }} \right]$ along the 
following procedure. (i) First, the approximate form of the effective pair 
potential which induces the OPSS with the spin-singlet or spin-triplet 
symmetry is proposed. (ii) Using such a potential, we derive the approximate 
solution of the BdG-KS equation given by Eqs. (\ref{eq53}) 
and (\ref{eq92}) -- (\ref{eq95}). The 
resultant solution of the energy spectrum possesses the energy gap for the 
excitation, which seems to be physically reasonable for the superconducting 
state. (iii) Finally, we derive the xc energy functional that yields the 
above-mentioned effective pair potential. It is shown that such an 
approximate form is related to the attractive interaction that is included 
in the Hamiltonian Eq. (\ref{eq5}). In what follows, we shall show the details.

\vspace{5mm}
\noindent
\textit{Step (i) }

Suppose that the effective pair potential that induces the OPSS with the 
spin symmetry $\tau $ is denoted as $D_{s,\tau } ({\rm {\bf r}}\zeta ,{\rm 
{\bf {r}'}}{\zeta }')$. Here the spin wave functions of the first, second 
and third terms of Eq. (\ref{eq3}), which are the spin-triplet wave functions, are 
labeled as $\tau_{1} ,\,\tau_{2} ,\,\tau_{3} $, respectively, and the 
spin wave function of the fourth term of Eq. (\ref{eq3}), which is the spin-singlet 
one, is labeled as $\tau_{4} $. $D_{s,\tau } ({\rm {\bf r}}\zeta ,{\rm {\bf 
{r}'}}{\zeta }')$ is assumed to be split into the spin-dependent and 
spatial-dependent parts as follows:
\begin{equation}
\label{eq113}
D_{s,\tau } ({\rm {\bf r}}\zeta ,{\rm {\bf {r}'}}{\zeta 
}')={\frac{1}{2}}D_{s1,\tau } (\zeta ,{\zeta }')D_{s2,\tau } ({\rm {\bf 
r}},{\rm {\bf {r}'}}).
\end{equation}
Let us consider the concrete forms of $D_{s1,\,\tau } (\zeta ,{\zeta }')$ 
and $D_{s2,\,\tau } ({\rm {\bf r}},{\rm {\bf {r}'}})$ which induce the OPSS 
in the reference system. The OPSS given by Eq. (\ref{eq3}) is rewritten in the 
following form:
\begin{equation}
\label{eq114}
\hat{{\Delta }}({\rm {\bf r}}\zeta ,{\rm {\bf {r}'}}{\zeta 
}')=\sum\limits_\tau {\sum\limits_{k_{i} } {\sum\limits_{k_{j} } {C_{k_{i} 
\sigma_{\tau } } C_{k_{j} {\sigma }'_{\tau } } \Psi_{\tau }^{k_{i} k_{j} } 
({\rm {\bf r}},{\rm {\bf {r}'}})\Phi_{\tau } (\zeta ,{\zeta }')} } } ,
\end{equation}
where $\Psi_{\tau }^{k_{i} k_{j} } ({\rm {\bf r}},{\rm {\bf {r}'}})$ and 
$\Phi_{\tau } (\zeta ,{\zeta }')$ are the spatial wave function and spin 
one for the OPSS with the spin symmetry $\tau $, respectively. Substituting 
Eqs. (\ref{eq113}) and (\ref{eq114}) into the third term of Eq. (\ref{eq43}), the integrals which 
are dependent on the spin wave function and spatial wave function are, 
respectively, expressed as
\begin{equation}
\label{eq115}
\int\!\!\!\int {D_{s1,\,\tau }^{\ast }(\zeta ,{\zeta }')\Phi_{\tau } 
(\zeta ,{\zeta }')d\zeta d{\zeta }'} ,
\end{equation}
and
\begin{equation}
\label{eq116}
\int\!\!\!\int {D_{s2,\,\tau }^{\ast }({\rm {\bf r}},{\rm {\bf {r}'}})\Psi 
_{\tau }^{k_{i} k_{j} } ({\rm {\bf r}},{\rm {\bf {r}'}})d^{3}rd^{3}{r}'} .
\end{equation}
As an illustrative form of $D_{s1,\,\tau } (\zeta ,{\zeta }')$ that gives 
the nonzero value of Eq. (\ref{eq115}), we present
\begin{equation}
\label{eq117}
D_{s1,\,\tau } (\zeta ,{\zeta }')=\left\{ {{\begin{array}{*{20}c}
 {{\displaystyle \frac{1}{2}}\left( {\delta_{\zeta {\zeta }'} +{\left\langle {\zeta } 
\right|}\sigma_{z} {\left| {\zeta} \right\rangle }} 
\right)\,\,\,\,\,\mbox{for}\,\,\,\tau =\tau_{1} ,} \hfill \vspace{2mm}\\
 {{\displaystyle \frac{1}{2}}\left( {\delta_{\zeta {\zeta }'} -{\left\langle {\zeta } 
\right|}\sigma_{z} {\left| {\zeta} \right\rangle }} 
\right)\,\,\,\,\mbox{for}\,\,\,\tau =\tau_{2} ,} \hfill \vspace{2mm}\\
 {{\left\langle {\zeta } \right|}\sigma_{x} {\left| {\zeta} \right\rangle 
}\,\,\,\,\,\,\,\,\,\,\,\,\,\,\,\,\,\,\,\,\,\,\,\,\,\,\,\,\,\,\,\mbox{for}\,\,\,\tau 
=\tau_{3} ,} \hfill \vspace{2mm} \\
 {{\left\langle {\zeta } \right|}\sigma_{y} {\left| {\zeta} \right\rangle 
}\,\,\,\,\,\,\,\,\,\,\,\,\,\,\,\,\,\,\,\,\,\,\,\,\,\,\,\,\,\,\,\mbox{for}\,\,\,\tau 
=\tau_{4} ,} \hfill \\
\end{array} }} \right.
\end{equation}
where $\sigma_{x} $, $\sigma_{y} $ and $\sigma_{z} $ are the Pauli 
marices. 

On the other hand, in order to make Eq. (\ref{eq116}) not be zero, the 
spatial-dependent part $D_{s2,\,\tau } ({\rm {\bf r}},{\rm {\bf {r}'}})$ 
needs to possess the following properties:
\begin{equation}
\label{eq118}
\begin{array}{l}
 D_{s2,\,\tau } ({\rm {\bf r}},{\rm {\bf {r}'}})-D_{s2,\,\tau } ({\rm {\bf 
{r}'}},{\rm {\bf r}})\ne 0\,\,\,\,\,\,\,\mbox{for}\,\,\,\,\tau =\tau_{1} 
,\,\tau_{2} ,\,\tau_{3} , \\ 
 D_{s2,\,\tau } ({\rm {\bf r}},{\rm {\bf {r}'}})+D_{s2,\,\tau } ({\rm {\bf 
{r}'}},{\rm {\bf r}})\ne 0\,\,\,\,\,\,\,\mbox{for}\,\,\,\,\tau =\tau_{4} , 
\\ 
 \end{array}
\end{equation}
because $\Psi_{\tau }^{k_{i} k_{j} } ({\rm {\bf r}},{\rm {\bf {r}'}})$ is 
the antisymmetric function with respect to the permutation of the spatial 
coordinates in the cases of $\tau =\tau_{1} ,\,\tau_{2} ,\,\tau_{3} $, 
while it is the symmetric function in the case of $\,\tau =\tau_{4} $. For 
the latter convenience, we define the function $d_{s2,\,\tau } ({\rm {\bf 
r}},{\rm {\bf {r}'}})$ as follows:
\begin{equation}
\label{eq119}
d_{s2,\,\tau } ({\rm {\bf r}},{\rm {\bf {r}'}})=\left\{ 
{{\begin{array}{*{20}c}
 {{\displaystyle\frac{1}{2}}\left\{ {D_{s2,\,\tau } ({\rm {\bf r}},{\rm {\bf 
{r}'}})-D_{s2,\,\tau } ({\rm {\bf {r}'}},{\rm {\bf r}})} 
\right\}\,\,\,\,\,\,\,\,\,\,\,\mbox{for}\,\,\,\,\tau_{1} ,\,\tau_{2} ,\,\tau_{3} ,} 
\hfill \ \ \ \ \vspace{2mm} \\ 
 {{\displaystyle \frac{1}{2}}\left\{ {D_{s2,\,\tau } ({\rm {\bf r}},{\rm {\bf 
{r}'}})+D_{s2,\,\tau } ({\rm {\bf {r}'}},{\rm {\bf r}})} 
\right\}\,\,\,\,\,\,\,\,\,\,\,\mbox{for}\,\,\,\,\tau =\tau_{4} .} \hfill \ \ \ \ \ \ \ \\
\end{array} }} \right.
\end{equation}
As can be seen in Eq. (\ref{eq53}), the effective pair potential is included in a 
form of Eq. (\ref{eq54}). Using Eq. (\ref{eq119}), and considering the symmetry of Eq. 
(\ref{eq117}), Eq. (\ref{eq54}) is rewritten as
\begin{equation}
\label{eq120}
\tilde{{D}}_{s,\tau } ({\rm {\bf r}}\zeta ,{\rm {\bf {r}'}}{\zeta 
}')=D_{s1,\tau } (\zeta ,{\zeta }')d_{s2,\tau } ({\rm {\bf r}},{\rm {\bf 
{r}'}}).
\end{equation}

\vspace{5mm}
\noindent
\textit{Step (ii)}

Using Eq. (\ref{eq120}), let us consider the approximate solution of Eq. (\ref{eq53}). We 
adopt the approximation method presented by P. G. de Gennes \cite{48}. 
Specifically, the solution of the normal state is modified to be suitable 
for the superconducting state \cite{48}. The solution of the normal state obeys 
the equation
\begin{equation}
\label{eq121}
\left( {h_{s}^{{\rm {\bf r}}} -\mu } \right)w_{i} ({\rm {\bf 
r}})=\varepsilon_{i} w_{i} ({\rm {\bf r}}),
\end{equation}
where $h_{s}^{{\rm {\bf r}}} $ is given by Eq. (\ref{eq52}). In the case of the 
spin-singlet ($\tau =\tau_{4} )$, the solution of Eq. (\ref{eq53}) is searched in 
the following form \cite{48}:
\begin{equation}
\label{eq122}
\begin{array}{l}
 u_{i} ({\rm {\bf r}}\zeta )=\bar{{u}}_{i} \chi_{u_{i} } (\zeta )w_{i} 
({\rm {\bf r}}), \\ 
 v_{i} ({\rm {\bf r}}\zeta )=\bar{{v}}_{i} \chi_{v_{i} } (\zeta )w_{i} 
({\rm {\bf r}}), \\ 
 \end{array}
\end{equation}
where $\bar{{u}}_{i} $ and $\bar{{v}}_{i} $ are complex numbers. In order to 
satisfy the orthonormality of the solutions of the BdG-KS equation (Sec. IV 
C), we suppose that
\begin{eqnarray}
\label{eq123}
&&\int {\chi_{u_{i} } (\zeta )\chi_{v_{i} } (\zeta )d\zeta =0} ,
\\
\label{eq124}
&&\left| {\bar{{u}}_{i} } \right|^{2}+\left| {\bar{{v}}_{i} } \right|^{2}=1.
\end{eqnarray}
These relations guarantee Eqs. (\ref{eq62}), (\ref{eq64}) and (\ref{eq65}) to hold. Substituting Eq. 
(\ref{eq122}) into Eq. (\ref{eq53}), we can easily obtain
\begin{equation}
\label{eq125}
\left( {{\begin{array}{*{20}c}
 {\varepsilon_{i} -E_{i} } \hfill & {\begin{array}{l}
 \int {\chi_{u_{i} } (\zeta )D_{s1,\tau_{4} } (\zeta ,{\zeta }')\chi 
_{v_{i} } ({\zeta }')d\zeta d{\zeta }'} \\ 
 \!\times\! \int {w_{i}^{\ast }({\rm {\bf r}})d_{s2,\tau_{4} } ({\rm {\bf 
r}},{\rm {\bf {r}'}})w_{i}^{\ast }({\rm {\bf {r}'}})d^{3}rd^{3}{r}'} \\ 
 \end{array}} \hfill \\
 {\begin{array}{l}
\!\! -\!\int {\chi_{v_{i} } (\zeta )D_{s1,\tau_{4} }^{\ast } (\zeta ,{\zeta 
}')\chi_{u_{i} } ({\zeta }')d\zeta d{\zeta }'} \\ 
 \times \int {w_{i} ({\rm {\bf r}})d_{s2,\tau_{4} }^{\ast }({\rm {\bf 
r}},{\rm {\bf {r}'}})w_{i} ({\rm {\bf {r}'}})d^{3}rd^{3}{r}'} \\ 
 \end{array}} \hfill & {-\left( {\varepsilon_{i} +E_{i} } \right)} \hfill 
\\
\end{array} }} \!\!\right)\!\!\left( {{\begin{array}{*{20}c}
 \!{\bar{{u}}_{i} } \!\hfill \\
 \!{\bar{{v}}_{i}^{\ast }} \!\hfill \\
\end{array} }} \right)\!\!=\!\!\left( {{\begin{array}{*{20}c}
 \!0\! \hfill \\
 \!0\! \hfill \\
\end{array} }} \right),
\end{equation}
which yields 
\begin{eqnarray}
\label{eq126}
E_{i}^{2}=\varepsilon_{i}^{2}&-&\int {\chi_{u_{i} } (\zeta )D_{s1,\tau 
_{4} } (\zeta ,{\zeta }')\chi_{v_{i} } (\zeta )d\zeta d{\zeta }'\int {\chi 
_{v_{i} } (\zeta )D_{s1,\tau_{4} }^{\ast } (\zeta ,{\zeta }')\chi_{u_{i} } 
(\zeta )d\zeta d{\zeta }'} } \nonumber \\ 
&& \times \left| {\int {w_{i}^{\ast }({\rm {\bf r}})d_{s2,\tau_{4} } ({\rm 
{\bf r}},{\rm {\bf {r}'}})w_{i}^{\ast }({\rm {\bf {r}'}})d^{3}rd^{3}{r}'} } 
\right|^{2}. 
\end{eqnarray}
Substituting Eq. (\ref{eq117}) into Eq. (\ref{eq126}), and if we use 
$\chi_{u_{i} } (\zeta )=\chi_{\uparrow } (\zeta ),\,\,\,\chi_{v_{i} } (\zeta )
=\chi_{\downarrow } (\zeta )$ 
as an example that is satisfied with Eq. (\ref{eq123}), then Eq. (\ref{eq126}) 
becomes
\begin{equation}
\label{eq127}
E_{i}^{2}=\varepsilon_{i}^{2}+\left| {D_{i}^{\mbox{\scriptsize {singlet}}} } 
\right|^{2}\,\,\,\,\,\mbox{with}\,\,\,\,\,D_{i}^{\mbox{\scriptsize {singlet}}} =\int 
{w_{i}^{\ast }({\rm {\bf r}})d_{s2,\tau_{4} } ({\rm {\bf r}},{\rm {\bf 
{r}'}})w_{i}^{\ast }({\rm {\bf {r}'}})d^{3}rd^{3}{r}'} .
\end{equation}
Since the energy gap appears in the energy spectrum of the superconducting 
state, the result Eq. (\ref{eq127}) seems to be physically reasonable \cite{48}.

Next we consider the case of the spin-triplet 
($\tau =\tau_{1} ,\,\tau_{2} ,\,\tau_{3} )$. 
Instead of Eq. (\ref{eq122}), we use 
\begin{equation}
\label{eq128}
\begin{array}{l}
 u_{i} ({\rm {\bf r}}\zeta )=\bar{{u}}_{i} \chi_{u_{i} } (\zeta )x_{i} 
({\rm {\bf r}}), \\ 
 v_{i} ({\rm {\bf r}}\zeta )=\bar{{v}}_{i} \chi_{v_{i} } (\zeta )y_{i} 
({\rm {\bf r}}), \\ 
 \end{array}
\end{equation}
where $x_{i} ({\rm {\bf r}})$ and $y_{i} ({\rm {\bf r}})$ are the solutions 
of Eq. (\ref{eq121}), and the corresponding energies $\kappa_{i} $ and $\lambda 
_{i} $ are different from each other, which implies the orthogonality 
between $x_{i} ({\rm {\bf r}})$ and $y_{i} ({\rm {\bf r}})$. This also 
guarantees Eqs. (\ref{eq62}), (\ref{eq64}) and (\ref{eq65}). 
Substituting Eq. (\ref{eq128}) into Eq. (\ref{eq53}), we get
\begin{equation}
\label{eq129}
\left( {{\begin{array}{*{20}c}
 {\kappa_{i} -E_{i} } \hfill & {\begin{array}{l}
 \int {\chi_{u_{i} } (\zeta )D_{s1,\tau } (\zeta ,{\zeta }')\chi_{v_{i} } 
({\zeta }')d\zeta d{\zeta }'} \\ 
 \times \int {x_{i}^{\ast }({\rm {\bf r}})d_{s2,\tau } ({\rm {\bf r}},{\rm 
{\bf {r}'}})y_{i}^{\ast }({\rm {\bf {r}'}})d^{3}rd^{3}{r}'} \\ 
 \end{array}} \hfill \\
 {\begin{array}{l}
\!-\!\int {\chi_{v_{i} } (\zeta )D_{s1,\tau }^{\ast } (\zeta ,{\zeta }')\chi 
_{u_{i} } ({\zeta }')d\zeta d{\zeta }'} \\ 
 \times \int {y_{i} ({\rm {\bf r}})d_{s2,\tau }^{\ast }({\rm {\bf r}},{\rm 
{\bf {r}'}})x_{i} ({\rm {\bf {r}'}})d^{3}rd^{3}{r}'} \\ 
 \end{array}} \hfill & {-\left( {\lambda_{i} +E_{i} } \right)} \hfill \\
\end{array} }} \right)\!\!\left( {{\begin{array}{*{20}c}
 \!{\bar{{u}}_{i} } \!\hfill \\
 \!{\bar{{v}}_{i}^{\ast }} \!\hfill \\
\end{array} }} \right)\!=\!\left( {{\begin{array}{*{20}c}
 \!0\! \hfill \\
 \!0\! \hfill \\
\end{array} }} \right),
\end{equation}
which immediately leads to
\begin{eqnarray}
\label{eq130}
 E_{i}^{2}+E_{i} \left( {\lambda_{i} -\kappa_{i} } \right)-\lambda_{i} 
\kappa_{i} \!\!
&-&
\!\!\int\! {\chi_{u_{i} } (\zeta )D_{s1,\tau } (\zeta ,{\zeta 
}')\chi_{v_{i} } ({\zeta }')d\zeta d{\zeta }'} 
\!\int\! {\chi_{v_{i} } (\zeta 
)D_{s1,\tau }^{\ast } (\zeta ,{\zeta }')\chi_{u_{i} } ({\zeta }')d\zeta 
d{\zeta }'} \nonumber \\ 
&&\times \left| {\int {y_{i}^{\ast }({\rm {\bf r}})d_{s2,\tau } ({\rm {\bf 
r}},{\rm {\bf {r}'}})x_{i}^{\ast }({\rm {\bf {r}'}})d^{3}rd^{3}{r}'} } 
\right|^{2}=0, 
\end{eqnarray}
where $\tau =\tau_{1} ,\,\tau_{2} ,\,\tau_{3} $, and Eq. (\ref{eq119}) is used. 
Hereafter, we shall deal with the case of $\tau =\tau_{1} $ as an example 
for the spin-triplet case. Substituting Eq. (\ref{eq117}) for the case of $\tau 
=\tau_{1} $ into Eq. (\ref{eq130}), and if we use $\chi_{u_{i} } (\zeta )=\chi 
_{\uparrow } (\zeta ),\,\,\,\chi_{v_{i} } (\zeta )=\chi_{\uparrow } (\zeta 
)$, then Eq. (\ref{eq130}) becomes
\begin{equation}
\label{eq131}
E_{i}^{2}+E_{i} \left( {\lambda_{i} \!-\!\kappa_{i} } \right)-\lambda_{i} 
\kappa_{i} -\left| {D_{i}^{\mbox{\scriptsize{triplet}}} } 
\right|^{2}\!=\!0 \,\,\,\,\,\,\,
\mbox{with}\,\,\,\,\,D_{i}^{\mbox{\scriptsize{triplet}}} \!=\!\int {y_{i}^{\ast 
}({\rm {\bf r}})d_{s2,\tau_{1} } ({\rm {\bf r}},{\rm {\bf {r}'}})x_{i} 
^{\ast }({\rm {\bf {r}'}})d^{3}rd^{3}{r}'} .\ \ 
\end{equation}
If we choose $x_{i} ({\rm {\bf r}})$ and $y_{i} ({\rm {\bf r}})$ such that 
the corresponding energies $\kappa_{i} $ and $\lambda_{i} $ are close to 
each other, i.e., $\kappa_{i} \approx \lambda_{i} $, then Eq. (\ref{eq131}) 
approximately becomes to
\begin{equation}
\label{eq132}
E_{i}^{2}\approx \kappa_{i}^{2}+
\left| {D_{i}^{\mbox{\scriptsize {triplet}}} } \right|^{2}.
\end{equation}
Similarly to the case of the spin-singlet, this result physically sounds 
reasonable due to the appearance of the energy gap. Note that choosing the 
normal states which satisfy $\kappa_{i} \ne \lambda_{i} $ but $\kappa_{i} 
\approx \lambda_{i} $ is easy in metals because the density of states near 
the Fermi level is generally high in metals.

\vspace{5mm}
\noindent
\textit{Step (iii)}

Thus, the approximate forms of the effective pair potential lead to the 
energy spectrum accompanied by the energy gap for excitations, i.e. Eqs. 
(\ref{eq127}) and (\ref{eq132}). Next let us propose the illustrative forms of $D_{s2,\tau } 
({\rm {\bf r}},{\rm {\bf {r}'}})$ (or $d_{s2,\,\tau } ({\rm {\bf r}},{\rm 
{\bf {r}'}}))$ appeared in Eqs. (\ref{eq127}) and (\ref{eq132}) with the aid of the spirit 
of the LDA of the DFT \cite{24}. After that, using these functions, we finally 
present the approximate expression for $F_{xc} \left[ {n,\,{\rm {\bf 
j}}_{p}^{(T)} ,\,\Delta ,\Delta^{\ast }} \right]$ that yields the effective 
pair potential $D_{s,\tau } ({\rm {\bf r}}\zeta ,{\rm {\bf {r}'}}{\zeta }')$ 
properly. 

The LDA of the DFT borrows the results of the exchange and correlation 
energies of the homogeneous electron liquid \cite{64}. The LDA expression for the 
xc energy functional is correct in the limiting system such as the 
homogeneous electron system. With reference to the manner of the LDA, we 
propose the illustrative form of the effective pair potential. Specifically, 
the effective pair potential is determined by requiring that the energy gap 
given by Eq. (\ref{eq127}) or (\ref{eq132}) coincides with the correct one in the limiting 
system where the results of the BCS theory are sufficiently satisfactory 
\cite{53}. If the actual system is not far from such a limiting system, then it 
is expected that the approximation mentioned above works well. This 
expectation is also the same as that of the LDA in the DFT \cite{24}. 

With this approximation the energy gap is identical with that of the BCS 
theory. In the case of the spin-singlet, we have
\begin{equation}
\label{eq133}
\Gamma_{\mbox{\scriptsize{BCS}}}^{2}=\left| {D_{i}^{\mbox{\scriptsize{singlet}}} } \right|^{2}=\left| {\int 
{w_{i}^{\ast }({\rm {\bf r}})d_{s2,\tau_{4} } ({\rm {\bf r}},{\rm {\bf 
{r}'}})w_{i}^{\ast }({\rm {\bf {r}'}})d^{3}rd^{3}{r}'} } \right|^{2},
\end{equation}
where $\Gamma_{BCS} $ 
is the energy gap of the BCS theory that is given by \cite{53}
\begin{equation}
\label{eq134}
\Gamma_{\mbox{\scriptsize{BCS}}} =2\hbar \omega_{D} e^{-{\frac{1}{V_{0} N(0)}}},
\end{equation}
where $\omega_{D} $ and $N(0)$ are the Debye frequency and the density of 
states at the Fermi level, respectively, and where $V_{0} $ is the 
attractive interaction made by the approximation such that $V_{{\rm {\bf 
k{k}'}}} $ appeared in Eq. (\ref{eq14}) is simplified into $-V_{0} $ for the 
electrons near the Fermi level \cite{53}. As a concrete form of $D_{s2,\tau_{4} 
} ({\rm {\bf r}},{\rm {\bf {r}'}})$ which meets Eq. (\ref{eq133}), we can choose
\begin{equation}
\label{eq135}
D_{s2,\tau_{4} } ({\rm {\bf r}},{\rm {\bf {r}'}})=w_{i} ({\rm {\bf 
r}})w_{i} ({\rm {\bf {r}'}})\left\{ {2\hbar \omega_{D} 
e^{-{\frac{1}{V_{0} N(0)}}}} \right\},
\end{equation}
where Eq. (\ref{eq119}) is used.

Next we consider the case of the spin-triplet ($\tau =\tau_{1} )$. It is 
assumed that the relation between the energy gap 
$\Gamma_{\mbox{\scriptsize{triplet}}} $ 
and attractive interaction $V_{0} $ is denoted as 
$\Gamma_{\mbox{\scriptsize{triplet}}} \left( {V_{0} } \right)$ 
instead of Eq. (\ref{eq134}). 
Similarly to Eq. (\ref{eq133}), we use the following approximation:
\begin{equation}
\label{eq136}
\Gamma_{\mbox{\scriptsize{triplet}}} \left( {V_{0} } \right)^{2}
=\left| {D_{i}^{\mbox{\scriptsize{triplet}}} } 
\right|^{2}=\left| {\int {y_{i}^{\ast }({\rm {\bf r}})d_{s2,\tau_{1} } 
({\rm {\bf r}},{\rm {\bf {r}'}})x_{i}^{\ast }({\rm {\bf 
{r}'}})d^{3}rd^{3}{r}'} } \right|^{2}.
\end{equation}
As a concrete form of $D_{s2,\tau_{1} } ({\rm {\bf r}},{\rm {\bf {r}'}})$ 
which meets Eq. (\ref{eq136}), we can choose
\begin{equation}
\label{eq137}
D_{s2,\,\tau } ({\rm {\bf r}},{\rm {\bf {r}'}})=\left\{ {y_{i} ({\rm {\bf 
r}})x_{i} ({\rm {\bf {r}'}})-y_{i} ({\rm {\bf {r}'}})x_{i} ({\rm {\bf r}})} 
\right\}\Gamma_{\mbox{\scriptsize{triplet}}} (V_{0} ),
\end{equation}
where the orthogonality between $x_{i} ({\rm {\bf r}})$ and $y_{i} ({\rm 
{\bf r}})$ and Eq. (\ref{eq119}) are used. Thus, the illustrative forms of 
$D_{s2,\tau_{1} } ({\rm {\bf r}},{\rm {\bf {r}'}})$ are obtained, i.e., Eq. 
(\ref{eq135}) for the spin-singlet case and Eq. (\ref{eq137}) for the spin-triplet case.

Finally let us consider the approximate expression for $F_{xc} \left[ 
{n,\,{\rm {\bf j}}_{p}^{(T)} ,\,\Delta ,\Delta^{\ast }} \right]$. Suppose 
that $F_{xc} \left[ {n,\,{\rm {\bf j}}_{p}^{(T)} ,\,\Delta ,\Delta^{\ast }} 
\right]$ is decomposed into two parts, i.e., the OPSS-dependent part 
$E_{xc}^{S} \left[ {n,\,{\rm {\bf j}}_{p}^{(T)} ,\,\Delta ,\Delta^{\ast }} 
\right]$ and the OPSS-independent part $E_{xc}^{N} \left[ {n,\,{\rm {\bf 
j}}_{p}^{(T)} } \right]$. Namely we have
\begin{equation}
\label{eq138}
F_{xc} \left[ {n,\,{\rm {\bf j}}_{p}^{(T)} ,\,\Delta ,\Delta^{\ast }} 
\right]=E_{xc}^{S} \left[ {n,\,{\rm {\bf j}}_{p}^{(T)} ,\,\Delta ,\Delta 
^{\ast }} \right]+E_{xc}^{N} \left[ {n,\,{\rm {\bf j}}_{p}^{(T)} } \right].
\end{equation}
Aiming at developing the approximate form that is effective for the small 
magnitude of the OPSS, we will take only the first-order term in the 
expansion of $E_{xc}^{S} \left[ {n,\,{\rm {\bf j}}_{p}^{(T)} ,\,\Delta 
,\Delta^{\ast }} \right]$ with respect to the OPSS. That is to say,
\begin{eqnarray}
\label{eq139}
E_{xc}^{S} \left[ {n,\,{\rm {\bf j}}_{p}^{(T)} ,\,\Delta ,\Delta^{\ast }} \right]
&=&
{\frac{1}{2}}\int {D_{s1,\tau } (\zeta ,{\zeta }')D_{s2,\tau } ({\rm 
{\bf r}},{\rm {\bf {r}'}})\Delta^{\ast }({\rm {\bf r}}\zeta ,{\rm {\bf 
{r}'}}{\zeta }')d^{3}rd\zeta d^{3}{r}'d{\zeta }'} \nonumber \\ 
&+&
{\frac{1}{2}}\int {D_{s1,\tau }^{\ast }(\zeta ,{\zeta }')D_{s2,\tau } 
^{\ast }({\rm {\bf r}},{\rm {\bf {r}'}})\Delta ({\rm {\bf r}}\zeta ,{\rm 
{\bf {r}'}}{\zeta }')d^{3}rd\zeta d^{3}{r}'d{\zeta }'} , 
\end{eqnarray}
where Eq. (\ref{eq139}) is consistently satisfied with 
Eqs. (\ref{eq94}) and (\ref{eq95}) if 
$D_{s2,\tau } ({\rm {\bf r}},{\rm {\bf {r}'}})$ is supposed to be 
independent of the OPSS. 
As can be seen in Eqs. (\ref{eq135}) and (\ref{eq137}), 
$D_{s2,\tau } ({\rm {\bf r}},{\rm {\bf {r}'}})$ 
is written by means of the wave 
functions of the normal state. Although they are formally dependent on the 
OPSS through the effective mean-field potentials $v_{s} ({\rm {\bf r}})$ and 
${\rm {\bf A}}_{s} ({\rm {\bf r}})$, the wave functions of the normal state 
may be approximately calculated by utilizing the xc energy functional which 
does not depend on the OPSS. This is because the electronic structures of 
the normal state have been usually and successfully calculated by means of 
the xc energy functional without the OPSS dependence \cite{65}.

When the effective pair potentials become zero, namely, when the system is 
in the normal state \cite{66}, Eq. (\ref{eq139}) becomes zero and the resulting $F_{xc} 
\left[ {n,\,{\rm {\bf j}}_{p}^{(T)} ,\,\Delta ,\Delta^{\ast }} \right]$ is 
equal to $E_{xc}^{N} \left[ {n,\,{\rm {\bf j}}_{p}^{(T)} } \right]$ in Eq. 
(\ref{eq138}). Therefore, $E_{xc}^{N} \left[ {n,\,{\rm {\bf j}}_{p}^{(T)} } \right]$ 
can be regarded as the xc energy functional of the normal state. As an 
approximate form, the vorticity expansion approximation (VEA) of the CDFT, 
which has been previously developed \cite{43,67,68,69,70}, would be available and 
useful. 

Substituting Eqs. (\ref{eq117}) and (\ref{eq135}) into Eq. (\ref{eq139}) and, for example, using the 
VEA expression \cite{43,67,68,69,70} as $E_{xc}^{N} \left[ {n,\,{\rm {\bf j}}_{p}^{(T)} 
} \right]$, the approximate form of $F_{xc} \left[ {n,\,{\rm {\bf 
j}}_{p}^{(T)} ,\,\Delta ,\Delta^{\ast }} \right]$ can be obtained for the 
case of the spin-singlet. On the other hand, substituting Eqs. (\ref{eq117}) and 
(\ref{eq137}) into Eq. (\ref{eq139}), the approximate form of the xc energy functional for 
the case of the spin-triplet ($\tau =\tau_{1} )$ can be obtained. It should 
be noted that the attractive interaction $V_{0} $ which is originally 
contained in the Hamiltonian Eq. (\ref{eq5}) is explicitly contained in $F_{xc} 
\left[ {n,\,{\rm {\bf j}}_{p}^{(T)} ,\,\Delta ,\Delta^{\ast }} \right]$. 
This is an advisable feature for the xc energy functional because it 
intrinsically includes the quantum effects of the electron-electron 
interaction. 
%
%*****************Sec. VII **********************************
\section{Concluding Remarks}
In this paper, we present the CDFT for the superconductor with the 
improvements on several issues listed as (i) -- (iv) in Sec. I. They are all 
indispensable for constructing the CDFT which is suitable for 
superconductors, though such improvements have not sufficiently been 
performed in the previous works \cite{26,27,45}. The present CDFT reproduces the 
equilibrium values of the OPSS, electron density and transverse component of 
the paramagnetic current-density for the superconductor immersed in the 
magnetic field. In what follows, we shall discuss the features of the 
present CDFT.

(A) As mentioned in Sec. II, the OPSS contains the information on the 
pairing states of the superconducting state. Since the OPSS is directly 
evaluated by the present CDFT, various kinds of superconducting properties 
can be discussed. 

(A-1) Not only the critical temperature but also the critical magnetic field 
can be predicted by evaluating the dependences of the OPSS on the 
temperature and on the external magnetic field. The temperature and magnetic 
field in the case when the OPSS disappears correspond to the critical 
temperature and critical magnetic field, respectively. 

(A-2) The spin symmetry for the pairing state is understood from the OPSS 
directly. It can be verified from Eq. (\ref{eq3}) whether the pairing state has as 
the spin part the spin-singlet symmetry or spin-triplet one or their mixed 
one.

(A-3) The spatial dependence of the OPSS explicitly shows the spatial 
symmetry and distribution of the pairing state. The dependence of the OPSS 
on the relative coordinates gives the spatial broadening of the individual 
pairing state. It clarifies to what extent the pairing state is close to the 
Bose particle. On the other hand, the dependence of the OPSS on the 
coordinates of the center of gravity gives the spatial distribution of the 
superconducting phase. The disappearing regions of the OPSS can be regarded 
as the spatial pattern of the magnetic vortex in the mixed state of the 
Type-II superconductor. 

(B) In addition to the OPSS, we can obtain the current-density in the 
present CDFT. This is a striking feature of this theory because the physical 
phenomena which are related to the current-density can be discussed 
directly. Specifically, we can discuss the Meissner effect, Silsbee's rule 
that gives the relation between the critical current-density and the 
critical magnetic field \cite{30,31}, and the energy gap of the superconductor. 
Before discussing these phenomena, we will explain the way to get the 
current-density (sum of the paramagnetic current-density and antimagnetic 
one) within the present theory. The present CDFT can predict the transverse 
component of the paramagnetic current-density as well as the OPSS and 
electron density. Using the equation of continuity and taking into account 
the Helmholtz theorem for the vector analysis \cite{71}, the longitudinal 
component of the paramagnetic current-density can be determined. Since the 
transverse component of the paramagnetic current-density is determined from 
the present CDFT directly, the whole component of the paramagnetic 
current-density can be obtained. On the other hand, the antimagnetic 
current-density can also be obtained from the present CDFT since it depends 
on the electron density that is one of the basic variables. Therefore, the 
sum of the paramagnetic current-density and antimagnetic one, i.e., 
current-density of the system, can be eventually obtained.

(B-1) The Meissner effect is generally investigated by calculating the 
magnetic field inside the superconducting system. The magnetic field does 
not exist in the superconducting phase except the boundary region where the 
current-density flows. The magnetic field is generally calculated from the 
current-density via the microscopic Maxwell equation \cite{59,60}. Since the 
current-density can be obtained from the present CDFT, as mentioned above, 
it can be verified whether the magnetic field in the superconducting phase 
becomes zero or not. The present CDFT enables us to make the direct 
observation of the Meissner effect.

(B-2) Silsbee's rule (or the London equation) provides the relation between 
the critical magnetic field and critical current-density \cite{30}. In the 
present CDFT, we can get the relation between the external magnetic field 
and the current-density existing when the OPSS disappears. That is to say, 
it can be argued to what extent the critical magnetic field and the 
corresponding current-density which are both calculated from the present 
CDFT are satisfied with Silsbee's rule \cite{72}. The present CDFT enables us to 
make the direct observation of Silsbee's rule as well as the Meissner 
effect.

(B-3) Besides the above phenomena, we can also discuss the energy gap of the 
superconductor by using the critical current-density. In general, the 
current-density flowing in the superconductor has the upper limit, and its 
magnitude is related to the energy gap. \cite{22} Using the critical 
current-density calculated from the present CDFT, we can evaluate the 
magnitude of the energy gap of the superconductor. Comparing this result 
with the approximate solution $\left| {D_{i}^{\mbox{\scriptsize{singlet}}} } \right|$ or 
$\left| {D_{i}^{\mbox{\scriptsize{triplet}}} } \right|$ given in Sec. VI, it is possible 
to discuss to what extent the system is close to the BCS case.

Thus, the present CDFT for the superconductor would be effective for 
investigating the properties of the superconducting state which are related 
to the magnetic field and/or current-density flowing in the superconductor.

\newpage
\appendix*
\section{Property of the BV transformation}

In this Appendix, we confirm that Eq. (\ref{eq44}) can be regarded as a 
transformation which is similar to the unitary transformation \cite{73}. Equation 
(\ref{eq44}) is collectively written in a form of using matrices as
\begin{equation}
\label{eq140}
\left( {{\begin{array}{*{20}c}
 {\psi ({\rm {\bf r}}_{1} \zeta_{1} )} \hfill \\
 {\psi^{\dag }({\rm {\bf r}}_{1} \zeta_{1} )} \hfill \\
 {\psi ({\rm {\bf r}}_{2} \zeta_{2} )} \hfill \\
 {\begin{array}{c}
 \psi^{\dag }({\rm {\bf r}}_{2} \zeta_{2} ) \\ 
 \vdots \\ 
 \vdots \\ 
 \end{array}} \hfill \\
\end{array} }} \right)=\left( {{\begin{array}{*{20}c}
 {u_{i_{1} } ({\rm {\bf r}}_{1} \zeta_{1} )} \hfill & {v_{i_{1} } ({\rm 
{\bf r}}_{1} \zeta_{1} )} \hfill & {u_{i_{2} } ({\rm {\bf r}}_{1} \zeta 
_{1} )} \hfill & {v_{i_{2} } ({\rm {\bf r}}_{1} \zeta_{1} )} \hfill & 
\cdots \hfill & \cdots \hfill \\
 {v_{i_{1} }^{\ast } ({\rm {\bf r}}_{1} \zeta_{1} )} \hfill & {u_{i_{1} 
}^{\ast } ({\rm {\bf r}}_{1} \zeta_{1} )} \hfill & {v_{i_{2} }^{\ast } 
({\rm {\bf r}}_{1} \zeta_{1} )} \hfill & {u_{i_{2} }^{\ast } ({\rm {\bf 
r}}_{1} \zeta_{1} )} \hfill & \cdots \hfill & \cdots \hfill \\
 {u_{i_{1} } ({\rm {\bf r}}_{2} \zeta_{2} )} \hfill & {v_{i_{1} } ({\rm 
{\bf r}}_{2} \zeta_{2} )} \hfill & {u_{i_{2} } ({\rm {\bf r}}_{2} \zeta 
_{2} )} \hfill & {v_{i_{2} } ({\rm {\bf r}}_{2} \zeta_{2} )} \hfill & 
\cdots \hfill & \cdots \hfill \\
 {v_{i_{1} }^{\ast } ({\rm {\bf r}}_{2} \zeta_{2} )} \hfill & {u_{i_{1} 
}^{\ast } ({\rm {\bf r}}_{2} \zeta_{2} )} \hfill & {v_{i_{2} }^{\ast } 
({\rm {\bf r}}_{2} \zeta_{2} )} \hfill & {u_{i_{2} }^{\ast } ({\rm {\bf 
r}}_{2} \zeta_{2} )} \hfill & \cdots \hfill & \cdots \hfill \\
 \vdots \hfill & \vdots \hfill & \vdots \hfill & \vdots \hfill & \ddots 
\hfill & \hfill \\
 \vdots \hfill & \vdots \hfill & \vdots \hfill & \vdots \hfill & \hfill & 
\ddots \hfill \\
\end{array} }} \right)\left( {{\begin{array}{*{20}c}
 {\gamma_{i_{1} } } \hfill \\
 {\gamma_{i_{1} }^{\dag } } \hfill \\
 {\gamma_{i_{2} } } \hfill \\
 {\gamma_{i_{2} }^{\dag } } \hfill \\
 \vdots \hfill \\
 \vdots \hfill \\
\end{array} }} \right),
\end{equation}
where ${\rm {\bf r}}_{i} $ and ${\rm {\bf r}}_{i+1} $ are, respectively, 
located in the infinitesimal volumes which are adjacent to each other. The 
infinitesimal volume is assumed to be $\delta^{3}$. Let us observe the 
property of the transformation matrix of Eq. (\ref{eq140}). If it is denoted as 
$\hat{{U}}$, then we have
\begin{equation}
\label{eq141}
\hat{{U}}=\left( {{\begin{array}{*{20}c}
 {\varpi_{i_{1} } ({\rm {\bf r}}_{1} \zeta_{1} )} \hfill & {\varpi_{i_{2} 
} ({\rm {\bf r}}_{1} \zeta_{1} )} \hfill & {\varpi_{i_{3} } ({\rm {\bf 
r}}_{1} \zeta_{1} )} \hfill & \cdots \hfill & \cdots \hfill \\
 {\varpi_{i_{1} } ({\rm {\bf r}}_{2} \zeta_{2} )} \hfill & {\varpi_{i_{2} 
} ({\rm {\bf r}}_{2} \zeta_{2} )} \hfill & {\varpi_{i_{3} } ({\rm {\bf 
r}}_{2} \zeta_{2} )} \hfill & \cdots \hfill & \cdots \hfill \\
 {\varpi_{i_{1} } ({\rm {\bf r}}_{3} \zeta_{3} )} \hfill & {\varpi_{i_{2} 
} ({\rm {\bf r}}_{3} \zeta_{3} )} \hfill & {\varpi_{i_{3} } ({\rm {\bf 
r}}_{3} \zeta_{3} )} \hfill & \cdots \hfill & \cdots \hfill \\
 \vdots \hfill & \vdots \hfill & \vdots \hfill & \ddots \hfill & \hfill \\
 \vdots \hfill & \vdots \hfill & \vdots \hfill & \hfill & \ddots \hfill \\
\end{array} }} \right),
\end{equation}
where $\varpi_{i_{\alpha } } ({\rm {\bf r}}_{\beta } \zeta_{\beta } )$ is 
the constituent $2\times 2$ block matrix defined as
\begin{equation}
\label{eq142}
\varpi_{i_{\alpha } } ({\rm {\bf r}}_{\beta } \zeta_{\beta } )=\left( 
{{\begin{array}{*{20}c}
 {u_{i_{\alpha } } ({\rm {\bf r}}_{\beta } \zeta_{\beta } )} \hfill & 
{v_{i_{\alpha } } ({\rm {\bf r}}_{\beta } \zeta_{\beta } )} \hfill \\
 {v_{i_{\alpha } }^{\ast } ({\rm {\bf r}}_{\beta } \zeta_{\beta } )} \hfill 
& {u_{i_{\alpha } }^{\ast } ({\rm {\bf r}}_{\beta } \zeta_{\beta } )} 
\hfill \\
\end{array} }} \right).
\end{equation}
We shall consider the product $\hat{{U}}^{\dag }\hat{{U}}$:
\begin{equation}
\label{eq143}
\hat{{U}}^{\dag }\hat{{U}}=\left( {{\begin{array}{*{20}c}
 {\sum\limits_{{\rm {\bf r}}} {\sum\limits_\zeta {\varpi_{i_{1} }^{\dag } 
({\rm {\bf r}}\zeta )\varpi_{i_{1} } ({\rm {\bf r}}\zeta )} } } \hfill & 
{\sum\limits_{{\rm {\bf r}}} {\sum\limits_\zeta {\varpi_{i_{1} }^{\dag } 
({\rm {\bf r}}\zeta )\varpi_{i_{2} } ({\rm {\bf r}}\zeta )} } } \hfill & 
{\sum\limits_{{\rm {\bf r}}} {\sum\limits_\zeta {\varpi_{i_{1} }^{\dag } 
({\rm {\bf r}}\zeta )\varpi_{i_{3} } ({\rm {\bf r}}\zeta )} } } \hfill & 
\cdots \hfill & \cdots \hfill \\
 {\sum\limits_{{\rm {\bf r}}} {\sum\limits_\zeta {\varpi_{i_{2} }^{\dag } 
({\rm {\bf r}}\zeta )\varpi_{i_{1} } ({\rm {\bf r}}\zeta )} } } \hfill & 
{\sum\limits_{{\rm {\bf r}}} {\sum\limits_\zeta {\varpi_{i_{2} }^{\dag } 
({\rm {\bf r}}\zeta )\varpi_{i_{2} } ({\rm {\bf r}}\zeta )} } } \hfill & 
{\sum\limits_{{\rm {\bf r}}} {\sum\limits_\zeta {\varpi_{i_{2} }^{\dag } 
({\rm {\bf r}}\zeta )\varpi_{i_{3} } ({\rm {\bf r}}\zeta )} } } \hfill & 
\cdots \hfill & \cdots \hfill \\
 {\sum\limits_{{\rm {\bf r}}} {\sum\limits_\zeta {\varpi_{i_{3} }^{\dag } 
({\rm {\bf r}}\zeta )\varpi_{i_{1} } ({\rm {\bf r}}\zeta )} } } \hfill & 
{\sum\limits_{{\rm {\bf r}}} {\sum\limits_\zeta {\varpi_{i_{3} }^{\dag } 
({\rm {\bf r}}\zeta )\varpi_{i_{2} } ({\rm {\bf r}}\zeta )} } } \hfill & 
{\sum\limits_{{\rm {\bf r}}} {\sum\limits_\zeta {\varpi_{i_{3} }^{\dag } 
({\rm {\bf r}}\zeta )\varpi_{i_{3} } ({\rm {\bf r}}\zeta )} } } \hfill & 
\cdots \hfill & \cdots \hfill \\
 \vdots \hfill & \vdots \hfill & \vdots \hfill & \ddots \hfill & \hfill \\
 \vdots \hfill & \vdots \hfill & \vdots \hfill & \hfill & \ddots \hfill \\
\end{array} }} \right).
\end{equation}
The diagonal $2\times 2$ block matrix is written as
\begin{equation}
\label{eq144}
\begin{array}{c}
 \sum\limits_{{\rm {\bf r}}} {\sum\limits_\zeta {\varpi_{i_{\alpha } 
}^{\dag } ({\rm {\bf r}}\zeta )\varpi_{i_{\alpha } } ({\rm {\bf r}}\zeta )} 
} =\sum\limits_{{\rm {\bf r}}} {\sum\limits_\zeta {\left( 
{{\begin{array}{*{20}c}
 {\begin{array}{l}
 u_{i_{\alpha } }^{\ast } ({\rm {\bf r}}\zeta )u_{i_{\alpha } } ({\rm {\bf 
r}}\zeta ) \\ 
 +v_{i_{\alpha } } ({\rm {\bf r}}\zeta )v_{i_{\alpha } }^{\ast } ({\rm {\bf 
r}}\zeta ) \\ 
 \end{array}} \hfill & {\begin{array}{l}
 u_{i_{\alpha } }^{\ast } ({\rm {\bf r}}\zeta )v_{i_{\alpha } } ({\rm {\bf 
r}}\zeta ) \\ 
 +v_{i_{\alpha } } ({\rm {\bf r}}\zeta )u_{i_{\alpha } }^{\ast } ({\rm {\bf 
r}}\zeta ) \\ 
 \end{array}} \hfill \\
 {\begin{array}{l}
 v_{i_{\alpha } }^{\ast } ({\rm {\bf r}}\zeta )u_{i_{\alpha } } ({\rm {\bf 
r}}\zeta ) \\ 
 +u_{i_{\alpha } } ({\rm {\bf r}}\zeta )v_{i_{\alpha } }^{\ast } ({\rm {\bf 
r}}\zeta ) \\ 
 \end{array}} \hfill & {\begin{array}{l}
 v_{i_{\alpha } }^{\ast } ({\rm {\bf r}}\zeta )v_{i_{\alpha } } ({\rm {\bf 
r}}\zeta ) \\ 
 +u_{i_{\alpha } } ({\rm {\bf r}}\zeta )u_{i_{\alpha } }^{\ast } ({\rm {\bf 
r}}\zeta ) \\ 
 \end{array}} \hfill \\
\end{array} }} \right)} } \\ 
 =\displaystyle{\frac{1}{\delta^{3}}}\left( {{\begin{array}{*{20}c}
 1 \hfill & 0 \hfill \\
 0 \hfill & 1 \hfill \\
\end{array} }} \right), \\ 
 \end{array}
\end{equation}
where we use Eqs. (\ref{eq62}), (\ref{eq65}), and the general relation 
$\sum\limits_{{\rm {\bf r}}_{i} } {f({\rm {\bf r}}_{i} )\delta^{3}} 
=\int {f({\rm {\bf r}})} d^{3}r$. 
On the other hand, the off-diagonal $2\times 2$ block matrix is 
written as
\begin{equation}
\label{eq145}
\begin{array}{c}
 \sum\limits_{{\rm {\bf r}}} {\sum\limits_\zeta {\varpi_{i_{\alpha } 
}^{\dag } ({\rm {\bf r}}\zeta )\varpi_{i_{\beta } } ({\rm {\bf r}}\zeta )} 
} =\sum\limits_{{\rm {\bf r}}} {\sum\limits_\zeta {\left( 
{{\begin{array}{*{20}c}
 {\begin{array}{l}
 u_{i_{\alpha } }^{\ast } ({\rm {\bf r}}\zeta )u_{i_{\beta } } ({\rm {\bf 
r}}\zeta ) \\ 
 +v_{i_{\alpha } } ({\rm {\bf r}}\zeta )v_{i_{\beta } }^{\ast } ({\rm {\bf 
r}}\zeta ) \\ 
 \end{array}} \hfill & {\begin{array}{l}
 u_{i_{\alpha } }^{\ast } ({\rm {\bf r}}\zeta )v_{i_{\beta } } ({\rm {\bf 
r}}\zeta ) \\ 
 +v_{i_{\alpha } } ({\rm {\bf r}}\zeta )u_{i_{\beta } }^{\ast } ({\rm {\bf 
r}}\zeta ) \\ 
 \end{array}} \hfill \\
 {\begin{array}{l}
 v_{i_{\alpha } }^{\ast } ({\rm {\bf r}}\zeta )u_{i_{\beta } } ({\rm {\bf 
r}}\zeta ) \\ 
 +u_{i_{\alpha } } ({\rm {\bf r}}\zeta )v_{i_{\beta } }^{\ast } ({\rm {\bf 
r}}\zeta ) \\ 
 \end{array}} \hfill & {\begin{array}{l}
 v_{i_{\alpha } }^{\ast } ({\rm {\bf r}}\zeta )v_{i_{\beta } } ({\rm {\bf 
r}}\zeta ) \\ 
 +u_{i_{\alpha } } ({\rm {\bf r}}\zeta )u_{i_{\beta } }^{\ast } ({\rm {\bf 
r}}\zeta ) \\ 
 \end{array}} \hfill \\
\end{array} }} \right)} } \\ 
 =\left( {{\begin{array}{*{20}c}
 0 \hfill & 0 \hfill \\
 0 \hfill & 0 \hfill \\
\end{array} }} \right), \\ 
 \end{array}
\end{equation}
where we use Eqs. (\ref{eq62}), (\ref{eq64}) and the general relation 
mentioned above. 
Thus, Eqs. (\ref{eq144}) and (\ref{eq145}) lead to
\begin{equation}
\label{eq146}
\hat{{U}}^{\dag }\hat{{U}}={\frac{1}{\delta^{3}}}\hat{{I}},
\end{equation}
where $\hat{{I}}$ is the unit matrix.

From Eq. (\ref{eq146}), it can be shown that
\begin{equation}
\label{eq147}
\hat{{U}}\hat{{U}}^{\dag }={\frac{1}{\delta^{3}}}\hat{{I}}.
\end{equation}
We shall show the relations that are derived from Eq. (\ref{eq147}). 
The product $\hat{{U}}\hat{{U}}^{\dag }$ is given by
\begin{equation}
\label{eq148}
\hat{{U}}\hat{{U}}^{\dag }=\left( {{\begin{array}{*{20}c}
 {\sum\limits_i {\varpi_{i} ({\rm {\bf r}}_{1} \zeta_{1} )\varpi 
_{i}^{\dag } ({\rm {\bf r}}_{1} \zeta_{1} )} } \hfill & {\sum\limits_i 
{\varpi_{i} ({\rm {\bf r}}_{1} \zeta_{1} )\varpi_{i}^{\dag } ({\rm {\bf 
r}}_{2} \zeta_{2} )} } \hfill & {\sum\limits_i {\varpi_{i} ({\rm {\bf 
r}}_{1} \zeta_{1} )\varpi_{i}^{\dag } ({\rm {\bf r}}_{3} \zeta_{3} )} } 
\hfill & \cdots \hfill & \cdots \hfill \\
 {\sum\limits_i {\varpi_{i} ({\rm {\bf r}}_{2} \zeta_{2} )\varpi 
_{i}^{\dag } ({\rm {\bf r}}_{1} \zeta_{1} )} } \hfill & {\sum\limits_i 
{\varpi_{i} ({\rm {\bf r}}_{2} \zeta_{2} )\varpi_{i}^{\dag } ({\rm {\bf 
r}}_{2} \zeta_{2} )} } \hfill & {\sum\limits_i {\varpi_{i} ({\rm {\bf 
r}}_{2} \zeta_{2} )\varpi_{i}^{\dag } ({\rm {\bf r}}_{3} \zeta_{3} )} } 
\hfill & \cdots \hfill & \cdots \hfill \\
 {\sum\limits_i {\varpi_{i} ({\rm {\bf r}}_{3} \zeta_{3} )\varpi 
_{i}^{\dag } ({\rm {\bf r}}_{1} \zeta_{1} )} } \hfill & {\sum\limits_i 
{\varpi_{i} ({\rm {\bf r}}_{3} \zeta_{3} )\varpi_{i}^{\dag } ({\rm {\bf 
r}}_{2} \zeta_{2} )} } \hfill & {\sum\limits_i {\varpi_{i} ({\rm {\bf 
r}}_{3} \zeta_{3} )\varpi_{i}^{\dag } ({\rm {\bf r}}_{3} \zeta_{3} )} } 
\hfill & \cdots \hfill & \cdots \hfill \\
 \vdots \hfill & \vdots \hfill & \vdots \hfill & \ddots \hfill & \hfill \\
 \vdots \hfill & \vdots \hfill & \vdots \hfill & \hfill & \ddots \hfill \\
\end{array} }} \right).
\end{equation}
The diagonal and off-diagonal $2\times 2$ block matrices are written as
\begin{equation}
\label{eq149}
\sum\limits_i {\varpi_{i} ({\rm {\bf r}}\zeta )\varpi_{i}^{\dag } ({\rm 
{\bf r}}\zeta )} =\sum\limits_i {\left( {{\begin{array}{*{20}c}
 {\begin{array}{l}
 u_{i} ({\rm {\bf r}}\zeta )u_{i}^{\ast } ({\rm {\bf r}}\zeta ) \\ 
 +v_{i} ({\rm {\bf r}}\zeta )v_{i}^{\ast } ({\rm {\bf r}}\zeta ) \\ 
 \end{array}} \hfill & {\begin{array}{l}
 u_{i} ({\rm {\bf r}}\zeta )v_{i} ({\rm {\bf r}}\zeta ) \\ 
 +v_{i} ({\rm {\bf r}}\zeta )u_{i} ({\rm {\bf r}}\zeta ) \\ 
 \end{array}} \hfill \\
 {\begin{array}{l}
 v_{i}^{\ast } ({\rm {\bf r}}\zeta )u_{i}^{\ast } ({\rm {\bf r}}\zeta ) \\ 
 +u_{i}^{\ast } ({\rm {\bf r}}\zeta )v_{i}^{\ast } ({\rm {\bf r}}\zeta ) \\ 
 \end{array}} \hfill & {\begin{array}{l}
 v_{i}^{\ast } ({\rm {\bf r}}\zeta )v_{i} ({\rm {\bf r}}\zeta ) \\ 
 +u_{i}^{\ast } ({\rm {\bf r}}\zeta )u_{i} ({\rm {\bf r}}\zeta ) \\ 
 \end{array}} \hfill \\
\end{array} }} \right)} ,
\end{equation}
and
\begin{equation}
\label{eq150}
\sum\limits_i {\varpi_{i} ({\rm {\bf r}}_{\alpha } \zeta_{\alpha } )\varpi 
_{i}^{\dag } ({\rm {\bf r}}_{\beta } \zeta_{\beta } )} =\sum\limits_i 
{\left( {{\begin{array}{*{20}c}
 {\begin{array}{l}
 u_{i} ({\rm {\bf r}}_{\alpha } \zeta_{\alpha } )u_{i}^{\ast } ({\rm {\bf 
r}}_{\beta } \zeta_{\beta } ) \\ 
 +v_{i} ({\rm {\bf r}}_{\alpha } \zeta_{\alpha } )v_{i}^{\ast } ({\rm {\bf 
r}}_{\beta } \zeta_{\beta } ) \\ 
 \end{array}} \hfill & {\begin{array}{l}
 u_{i} ({\rm {\bf r}}_{\alpha } \zeta_{\alpha } )v_{i} ({\rm {\bf 
r}}_{\beta } \zeta_{\beta } ) \\ 
 +v_{i} ({\rm {\bf r}}_{\alpha } \zeta_{\alpha } )u_{i} ({\rm {\bf 
r}}_{\beta } \zeta_{\beta } ) \\ 
 \end{array}} \hfill \\
 {\begin{array}{l}
 v_{i}^{\ast } ({\rm {\bf r}}_{\alpha } \zeta_{\alpha } )u_{i}^{\ast } 
({\rm {\bf r}}_{\beta } \zeta_{\beta } ) \\ 
 +u_{i}^{\ast } ({\rm {\bf r}}_{\alpha } \zeta_{\alpha } )v_{i}^{\ast } 
({\rm {\bf r}}_{\beta } \zeta_{\beta } ) \\ 
 \end{array}} \hfill & {\begin{array}{l}
 v_{i}^{\ast } ({\rm {\bf r}}_{\alpha } \zeta_{\alpha } )v_{i} ({\rm {\bf 
r}}_{\beta } \zeta_{\beta } ) \\ 
 +u_{i}^{\ast } ({\rm {\bf r}}_{\alpha } \zeta_{\alpha } )u_{i} ({\rm {\bf 
r}}_{\beta } \zeta_{\beta } ) \\ 
 \end{array}} \hfill \\
\end{array} }} \right)} ,
\end{equation}
respectively. Compared these elements with Eq. (\ref{eq147}), we have
\begin{equation}
\label{eq151}
\begin{array}{l}
 \sum\limits_i {\left\{ {u_{i} ({\rm {\bf r}}_{\alpha } \zeta_{\alpha } 
)u_{i}^{\ast } ({\rm {\bf r}}_{\beta } \zeta_{\beta } )+v_{i} ({\rm {\bf 
r}}_{\alpha } \zeta_{\alpha } )v_{i}^{\ast } ({\rm {\bf r}}_{\beta } \zeta 
_{\beta } )} \right\}} =\displaystyle{\frac{1}{\delta^{3}}}\delta_{{\rm {\bf 
r}}_{\alpha } {\rm {\bf r}}_{\beta } } \delta_{\zeta_{\alpha } \zeta 
_{\beta } } , \\ 
 \sum\limits_i {\left\{ {u_{i} ({\rm {\bf r}}_{\alpha } \zeta_{\alpha } 
)v_{i} ({\rm {\bf r}}_{\beta } \zeta_{\beta } )+v_{i} ({\rm {\bf 
r}}_{\alpha } \zeta_{\alpha } )u_{i} ({\rm {\bf r}}_{\beta } \zeta_{\beta 
} )} \right\}=0} . \\ 
 \end{array}
\end{equation}
Using the general relation $\displaystyle{\frac{1}{\delta^{3}}}\delta_{{\rm {\bf 
r}}_{\alpha } {\rm {\bf r}}_{\beta } } \delta_{\zeta_{\alpha } \zeta 
_{\beta } } =\delta ({\rm {\bf r}}_{\alpha } -{\rm {\bf r}}_{\beta } )\delta 
_{\zeta_{\alpha } \zeta_{\beta } } $, Eq. (\ref{eq151}) exactly coincides with 
Eqs. (\ref{eq45}) and (\ref{eq46}). Thus, it is confirmed that the BV transformation owns 
the unitary-like property given by Eqs. (\ref{eq146}) and (\ref{eq147}).
\begin{acknowledgments}
This work was partially supported by Grant-in-Aid for Scientific Research 
(No. 26400354 and No. 26400397) of Japan Society for the Promotion of 
Science.
\end{acknowledgments}
\newpage


\begin{thebibliography}{00}
\bibitem{49}
This type of eigenfunction of the RDM2 is called type (a) which was 
defined in our recent paper cited by Ref. 50.
\bibitem{61}
The current-density included in the microscopic Maxwell equation is the 
sum of the paramagnetic current-density and antimagnetic current-density. 
Concerning the paramagnetic current-density, the transverse component is 
obtained from the present CDFT directly, and the longitudinal component is 
calculated by using both the equation of continuity and Helmholtz theorem 
\cite{71}. The antimagnetic current-density is also obtained from the present 
CDFT, because it depends on the electron density which is one of basic 
variables of this theory. The details will be discussed in Sec. VII.
\bibitem{63}
A similar expression can be found in the case of the conventional DFT 
at zero temperature \cite{23,24}, though the entropy terms do not appear in it. 
See, for example, Chap. 7 in Ref. 44.
\bibitem{66}
As can be seen in Eq. (\ref{eq135}), the case when the effective pair potential 
disappears corresponds to $V_{0} =0$ which means the nonexistence of the 
attractive interaction in the Hamiltonian Eq. (\ref{eq5}). Then, the system is 
necessarily in the normal state.
\bibitem{73}
In some textbooks, e.g., Ref. 48, this transformation is explicitly 
called the unitary transformation.
\end{thebibliography}
\end{document}